\documentclass[10pt,superscriptaddress,secnumarabic,amssymb,amsmath,nobibnotes,aps,prd,showkeys,showpacs,twocolumn,nofootinbib,floatfix]{revtex4-1}%

\usepackage[T1]{fontenc}
\usepackage[utf8]{inputenc}
\usepackage{bm}
\usepackage{mathrsfs}
\usepackage[all]{xy}
\usepackage{epsfig,subfigure}
\usepackage{amssymb, amsmath, amsfonts}
\usepackage[english]{babel}
\usepackage{makeidx}
\usepackage{mathrsfs}
\usepackage[colorlinks=true,breaklinks=true]{hyperref}
\usepackage{graphicx, wrapfig, epsf}
\usepackage{float}
\usepackage{comment}
\usepackage{enumitem}
\usepackage[normalem]{ulem}
\usepackage{cases}
\usepackage{float}
\usepackage{soul} 
\usepackage{booktabs} 
\usepackage[dvipsnames, table]{xcolor}
\hypersetup{urlcolor=BlueViolet,
	        citecolor=Plum,
	        linkcolor=OliveGreen}

\definecolor{Gray}{rgb}{.9,.9,.9}

\newcommand{\dd}{\mathrm{d}}

\newcommand{\solver}{\texttt{solver1d}}
\newcommand{\solverb}{\texttt{solver2d}}
\newcommand{\gwlens}{\texttt{GWLENS}}
\newcommand{\glow}{\texttt{GLoW}}
\newcommand{\glowsi}{\texttt{SingleIntegral}}
\newcommand{\glowmc}{\texttt{MultiContour}}
\newcommand{\bs}[1]{\boldsymbol{#1}}

\begin{document}

\title[]{A twist in the tail: \texttt{GWLENS}, a fast and accurate tool for wave-optics lensing of gravitational waves}

\author{Sreekanth Harikumar}
\email{sreekanth@camk.edu.pl}
\affiliation{Nicolaus Copernicus Astronomical Center, Polish Academy of Sciences, Bartycka 18, 00-716 Warsaw, Poland}

\author{Ansal Kanu}
\affiliation{Department of Physics, National Institute of Technology Silchar, Assam 788010, India}

\author{Anirudh Nemmani}
\affiliation{Nicolaus Copernicus Astronomical Center, Polish Academy of Sciences, Bartycka 18, 00-716 Warsaw, Poland}

\author{Pawe{\l} Cieciel{\k a}g}
\affiliation{Nicolaus Copernicus Astronomical Center, Polish Academy of Sciences, Bartycka 18, 00-716 Warsaw, Poland}

\author{Micha{\l} Bejger}
\affiliation{Nicolaus Copernicus Astronomical Center, Polish Academy of Sciences, Bartycka 18, 00-716 Warsaw, Poland}
\affiliation{INFN Sezione di Ferrara, Via Saragat 1, 44122 Ferrara, Italy}

\begin{abstract}
Gravitational-wave lensing in the wave-optics regime requires evaluating the Fresnel--Kirchhoff diffraction integral, whose integrand is highly oscillatory and must be evaluated numerically, as closed-form solutions exist only for a few lens models. We present a novel, efficient algorithm that splits the integration domain at a point beyond the region containing the lensed images. The bounded head region is integrated along the real axis with Gauss--Legendre quadrature, while the semi-infinite tail is rotated into the complex plane, where the integrand decays exponentially, and is evaluated with Gauss--Laguerre quadrature. Since the image region is never deformed, the method avoids  the treatment of poles and branch cuts required by existing contour-deformation approaches. Our results agree with the exact point-mass solution to $10^{-13}$ and show good agreement with other methods. We present several examples for non-trivial lens models, showing potential for parameter estimation of realistic lenses. The method is particularly efficient when the amplification factor is needed for many source positions. In addition to transient broadband compact binary coalescence
GW signals, we apply it to long-duration (continuous),
quasi-monochromatic GW signals, for which the relative motion
of source and lens makes the amplification factor time-dependent over the observation. The implementation is publicly available in the Python package \texttt{GWLENS}. 
\end{abstract}
\maketitle

\section{Introduction}

The propagation of gravitational waves (GWs), such as the compact binary coalescence (CBC) signals, is affected by the matter distribution along the line of sight, resulting in the phenomenon of GW lensing \cite{Takahashi2003,Takahashi2004thesis}. In contrast to electromagnetic waves, for which the geometric-optics approximation is valid over a wide range of astrophysical phenomena, the study of lensed GWs requires both the wave-optics and geometric-optics descriptions. In the regime where wave-optics effects become important \cite{Meena:2019ate,Mishra:2023ddt, Chan:2026alt, Bulashenko:2021fes}, the waveform undergoes a frequency-dependent modulation of the amplitude and phase. The LIGO-Virgo-KAGRA (LVK) Collaborations \cite{LIGOScientific:2021izm,LIGOScientific:2023bwz,LIGOScientific:2025cwb} and independent analyses   \cite{Janquart:2023mvf,Li:2019osa,Hannuksela:2019kle,McIsaac:2019use,Dai:2020tpj} carried out searches for lensed signals, finding no confident detection yet. The amplification factor needed for these searches was obtained by solving the GW propagation equation in the background geometry of the lens in the weak-field limit \cite{Nakamura_1999}, which yields the Fresnel--Kirchhoff diffraction integral \cite{Schneider:1992bmb}. 
Analytical solutions exist for a few lens models \cite{Deguchi1986,Matsunaga:2006uc}; however, for more complex lens models the integral has to be evaluated numerically.

The diffraction integral is highly oscillatory, so standard numerical integration techniques often fail to converge to the correct result. Several specialized methods for such oscillatory integrals have been proposed so far in the literature, including asymptotic series expansions \cite{Takahashi2004thesis,press1992numerical}, contour integration \cite{Ulmer:1994ij,Villarrubia-Rojo:2024xcj,Meena:2025gry,Mishra:2021xzz, Shan:2022xfx}, histogram counting \cite{Diego:2019lcd,Deka:2024ecp}, Levin-type methods \cite{Moylan:2007fi,Guo:2020eqw}, Filon quadrature \cite{Xiang2007}, $\epsilon$-regularization (also called adiabatic regularization), and Picard--Lefschetz theory, in which the integration domain over the lens plane is mapped into the complex plane \cite{Feldbrugge:2019fjs,Jow:2022pux}; for details of the convergence criteria and a comparison of these methods, we refer the reader to \cite{Guo:2020eqw, Tambalo:2022plm}. Some of these techniques are implemented in publicly available packages. For example, \texttt{Gravelamps} \cite{Wright:2021cbn} uses the Fourier integral technique for axisymmetric lenses, \texttt{GLoW} \cite{Villarrubia-Rojo:2024xcj,VillarrubiaRojo2025GLoWSoftware} and \texttt{Glworia} \cite{Cheung:2024ugg} use contour integration, and \texttt{FIONA} \cite{Ephremidze:2026era} uses non-uniform fast Fourier transforms and windowing. Notably, the \texttt{GLoW} implementation is fast for axisymmetric lenses, but slows down as the number of source positions required for long-duration signals increases; \texttt{FIONA} is fast for many source positions, but its performance degrades with increasing number of frequencies required for parameter estimation (PE) of the CBC signals.

In this work, we introduce the \textit{twisted-tail method}, a new technique for fast and accurate evaluation of the highly oscillatory integral that defines the lensing amplification. The method splits the integration domain into two regions, a \textit{head} and a \textit{tail}. The splitting point is chosen such that the stationary points of the time-delay surface, which correspond to the geometric-optics images, lie in the head. In the tail region, the integration contour is rotated into the complex plane, where the integrand decays exponentially. As this method is exact, it avoids the truncation errors inherent in series-expansion methods \cite{Guo:2020eqw}. We implement the method in the Python package\footnote{{\gwlens} repository: \url{https://github.com/hsreekanth15/GWLENS}\\Documentation: \url{https://hsreekanth15.github.io/GWLENS}} {\gwlens}: \texttt{G}ravitational \texttt{W}ave \texttt{L}ensing \texttt{E}xploration of \texttt{N}on-visible \texttt{S}tructures. Key features of {\gwlens} are as follows: 
\begin{itemize}[itemsep=1pt, topsep=1pt, parsep=1.5pt]
    \item Fast and accurate computation of diffraction integral in the wave-optics regime, 
    \item Axisymmetric and non-axisymmetric solvers (\texttt{solver1d} and \texttt{solver2d}),  
    \item CBC lensed signal PE with the command-line tool \texttt{gwlens\_pipe},  
    \item A waveform generator that supports direct use of the solver and optional lookup table for expensive lens models (class \texttt{LensedWaveformGenerator}). 
    \item Long-duration (``continuous waves'', CW) lensed signal  generation with the Doppler modulation and Solar System ephemerides (command-line interface \texttt{gwlens\_cw\_datagen}), 
    \item Support for user-defined arbitrary lens configurations,  
    \item Controllable optional parameters for speed and accuracy.    
\end{itemize}
 
The paper is organized as follows. We briefly introduce the theory of lensing in Sec.~\ref{sec:fk_diff_int}, and explain the twisted-tail method and its numerical implementation in \texttt{GWLENS} in Sec.~\ref{sec:twisted-tail}. The code structure of our pipeline is elaborated in Sec.~\ref{sec:code-structure}. Benchmarks, accuracy of the code and its comparison with other pipelines are presented in Sec.~\ref{sec:benchmarking}. We then discuss the applications to lensed CBC signal PE and lensing of CWs in Sec.~\ref{sec:applications}, and conclude in Sec.~\ref{sec:conclusion}.

\section{Fresnel--Kirchhoff diffraction integral}
\label{sec:fk_diff_int}

Our aim is to solve the highly oscillatory diffraction integral in the wave-optics regime where traditional numerical methods fail to converge. Readers who are interested in the derivation of this integral are referred to \cite{Schneider:1992bmb,Nakamura_1999,Takahashi2004thesis} and, for its application to GW lensing, to \cite{Harikumar:2026pnu, Jana:2024uta,Fairbairn:2022xln, Ubach:2025oxr,Seo:2026eto,Seo:2023rjd,Liao:2017ioi,Narola:2023viz,Singh:2025uvp,Grespan2023}. The Fresnel--Kirchhoff diffraction integral  or the lensing amplification factor in dimensionless form is given by 
\begin{equation} \label{eq:Fw}
    F(w, \bs{y}) = \frac{w}{2 \pi i} \int d^2 \bs{x}  e^{  i w T(\bs x,\bs y)}\,.
\end{equation}
The integration is performed over the dimensionless lens coordinates $\bs {x} = (\bs x_1 , \bs x_2)$ which are rescaled  by the characteristic length scale of the lens $\xi_0$ to make it dimensionless. The variable $\bs y$ is the dimensionless impact parameter which is obtained by rescaling with $\eta_0 = D_S \xi_0 /D_L$, where $D_L$ and $D_S$ are the angular diameter distances to the lens and source respectively. The quantity $w$ is called the dimensionless frequency which is proportional to the effective redshifted lens mass $M_{Lz}$ and the source frequency $f$ as 
\begin{equation}
    w = \frac{8 \pi G M_{Lz}}{c^3}f\,.
    \label{eq:w}
\end{equation}
The integral depends on the Fermat potential $T(\bs x, \bs y)$ which has contributions from the geometric and Shapiro time delays
\begin{equation}\label{fermat_potential}
    T(\bs x, \bs y) =  \frac{1}{2} | \bs x - \bs y|^2 - \psi(\bs x) + \phi_{m}(\bs y)\,,
\end{equation}
where $\psi(\bs x)$ is the effective lensing potential of the lens mass distribution projected onto the lens plane,  and  $\phi_{m}(\bs y)$ is a constant chosen such that $\mathrm{min}_{\bs x}T(\bs x,\bs y)= 0$. The stationarity condition $\nabla T = 0$ is the lens equation, whose solutions give the position of images in the geometric-optics limit. In this limit, the diffraction integral can be approximated using the method of stationary phase \cite{Takahashi2004thesis} provided the phase varies rapidly. However, in the wave-optics regime, where the wavelength is comparable to the characteristic lensing scale, the geometric-optics approximation breaks down and Eq.~\eqref{eq:Fw} has to be solved numerically. Developing such a numerical approach is the main objective of this work.  

\section{Twisted tail method}\label{sec:twisted-tail}

The method splits the radial part of the integral at a point $u_b$ as in the Fourier integral technique, with the split chosen beyond all images. The semi-infinite part (the \textit{tail}) is rotated into the complex plane, where the integrand decays exponentially. The finite part (the \textit{head}) is evaluated on the real axis with composite Gauss--Legendre quadrature and the tail region is evaluated with a scaled Gauss--Laguerre rule \cite{CHEN2012636,XU2016117,Milovanovic2017,Huybrechs2015}. The result is exact up to quadrature error and requires only the potential $\psi$. In this section, we introduce the general scheme of solving the diffraction integral in {\solverb}, applicable to non-axisymmetric lens configurations and then explain {\solver} as a special case applicable to axially-symmetric lens models. Both implementations allow users to write and supply arbitrary lensing potentials and obtain the lensing amplification factor.
\begin{figure}
    \centering
    \includegraphics[width=1.0\linewidth]{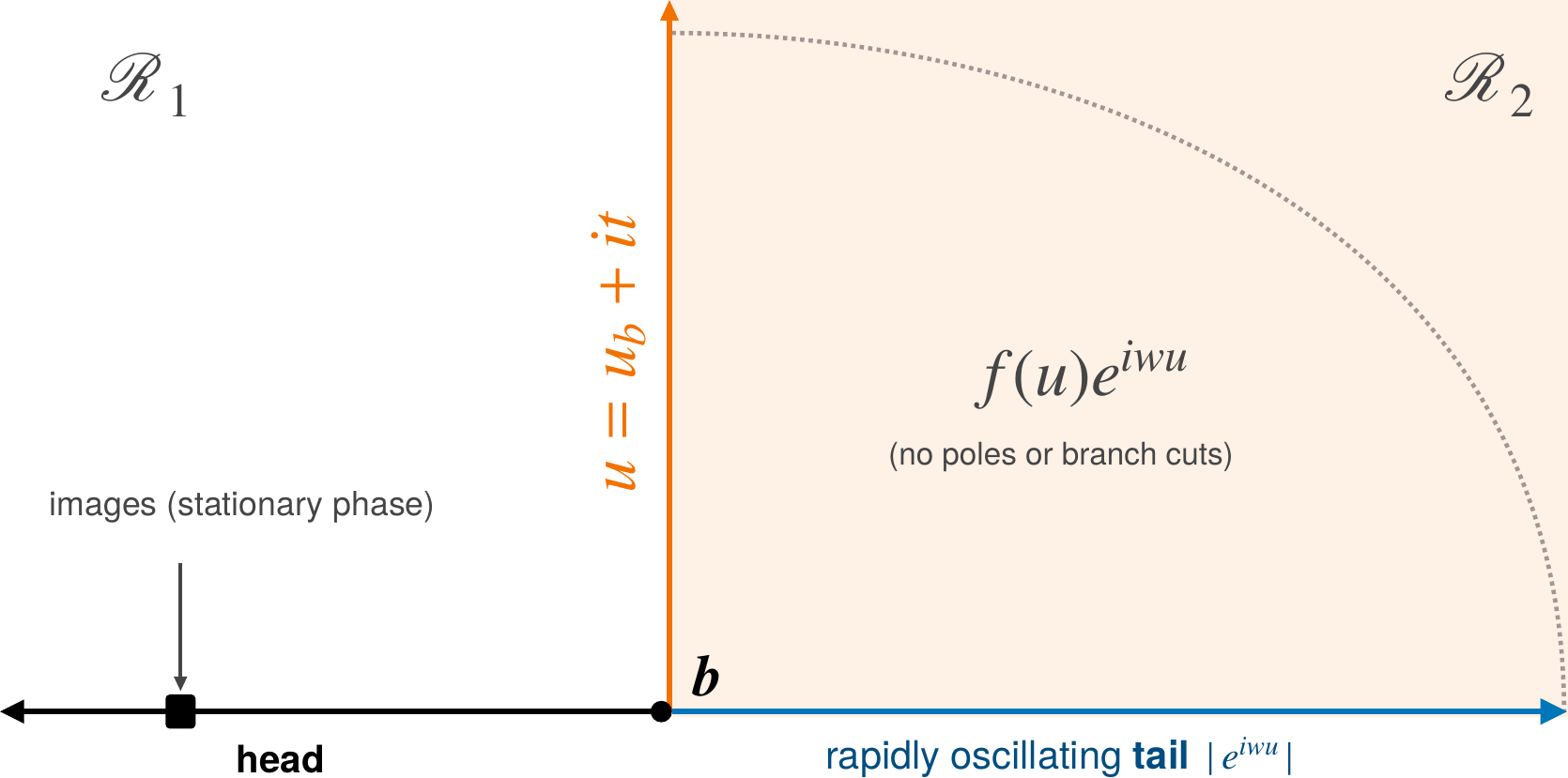}
    \caption{Graphical illustration of the twisted-tail method for an axisymmetric lens. The integration path is split at $u_b$ into the \textit{head} region $\mathcal{R}_1$, which contains the images and the lens-dominated oscillations and is integrated along the real axis, and to the highly oscillatory \textit{tail} region $\mathcal{R}_2$, which is rotated into the complex plane, where the integrand decays exponentially and can be evaluated efficiently. Cauchy's theorem guarantees that this contour deformation leaves the value of the integral unchanged. See Sec.~\ref{sec:twisted-tail} for details.}
    \label{fig:rtm-illustration}
\end{figure}

\subsection{General method}\label{subsec:general_method}

The algorithm starts by choosing a centre $\bf{c}$ which aligns with the singular points of one of the lenses in the lens plane. In polar coordinates about this centre,  $\bs{x} =  \bs{c} + r \hat{\bs{n}}(\theta)$ , where $\bs{n} = (\cos\theta, \sin \theta)$ is a unit vector $\hat{\bf{n}}.\hat{\bf{n}} = 1$, with ``.'' denoting the dot product. The position of the source measured with respect to the same centre is given by $\bf{b} = \bf{y} - \bf{c} $.
Transforming the radial coordinate using $u = r^2/2 $ as in the Fourier integral method \cite{Takahashi2004thesis}, the area element becomes $d^2 {\bs{x}} =  r dr\ d\theta = du\  d\theta$ and
\begin{equation}\label{eq:FW_2d_main}
 F = \frac{w}{2 \pi i} \int_0^{2\pi} d\theta \int_0^{\infty} du e^{i w T(u,\theta)}\,. 
\end{equation}
with the Fermat potential
\begin{equation}
    T(u,\theta) = u - \sqrt{2u}\;\hat{\bs{n}}\cdot\bs{b} + \frac{1}{2}|\bs{b}|^2 - \psi\big(\bs{c} + \sqrt{2 u}\,\hat{\bs{n}}\big)\,.
\end{equation}
The advantages of the above transformation are that no Jacobian factor remains and the dominant term of the phase $u = r^2/2 $ becomes linear in $u$; hence, far from the lens, the integrand oscillates at a constant rate.

The heart of the method is splitting the indefinite part of the integral into the  \textit{head}: $[0,u_b]$ and \textit{tail}: $[u_b, \infty)$. The head region is along the real axis while the tail part is moved onto the complex plane along the vertical line $u = u_b + i t$ as illustrated in Fig. \ref{fig:rtm-illustration}. The exactness of both the path along the horizontal direction (real)  and the vertical direction (imaginary) is guaranteed by Cauchy's integral theorem. Consider the closed contour that runs along the real axis from $u_b$ to $u_b + R$, then the quarter circle of radius $R$ centred on $u_b$, and back down the vertical line to $u_b$. All singular points of $\psi$ lie in the head so $e^{iwT}$ is analytic inside this closed contour and the loop integral vanishes \cite{boas2006mathematical}. On the real axis the tail integral converges only conditionally; as usual, $F$ is defined with a convergence factor $e^{-\epsilon u}$, $\epsilon\to 0^+$. With this factor the contribution of the arc vanishes as $R\to\infty$ (Jordan's lemma), and the limit $\epsilon\to 0$ can then be taken on the vertical path, where the integrand decays exponentially by itself (see below). The rotation is therefore exact, and it also turns a conditionally convergent oscillatory integral into an absolutely convergent, exponentially decaying one.

The head is the disc $|\bs{x}-\bs{c}| \le r_b$, which contains all images. To place the quadrature nodes, we first estimate how fast the integrand oscillates: the phase per unit $w$ accumulated from the centre, taken in the worst direction and including the largest source offset. The interval $[0,u_b]$ is then divided into sub-intervals $[a_m,b_m]$, each spanning a fixed set of phase cycles. On each sub-interval an  $n$-point Gauss--Legendre rule\footnote{A Gauss--Legendre integration rule is a numerical method of approximating a definite integral as a weighted sum of function values at specific points.} mapped from the interval $[-1,1]$ to $[a_m,b_m]$,
\begin{equation}
\begin{split}
u_{mk}&=\frac{a_m+b_m}{2}+\frac{b_m-a_m}{2}\,g_k\,,\\
W_{mk}&=\frac{b_m-a_m}{2}\,\omega_k\,,
\end{split}
\end{equation}
where $g_k$ are the roots of the Legendre polynomial $P_n$ and $\omega_k$ the corresponding weights.\footnote{The default value is \texttt{ngl = 32} Gauss--Legendre nodes computed with \texttt{scipy.special.roots\_legendre}.} The angular integral uses the $N$-point trapezoid rule, with nodes $\theta_l=2\pi l/N$, which converges exponentially for periodic integrands once $N$ exceeds the angular bandwidth of the integrand, $N \gtrsim w_{\max}\,(r\,|\bs{b}| + \max|\partial_\theta\psi|)$. Each node $\bs{x}_j=\bs{c}+r_{mk}\hat{\bs{n}}(\theta_l)$, with $r_{mk} = \sqrt{2u_{mk}}$, then carries the weight $W_j=W_{mk}\,(2\pi/N)\,\chi(\bs{x}_j)$. Here $\chi$ is a partition of unity.\footnote{A partition of unity is a set of smooth functions $\{\chi_k\}$ with $\chi_k(\bs{x})\geq 0$ and $\sum_k \chi_k(\bs{x})=1$ for every $\bs{x}$. Splitting $\int f\,\dd^2x=\sum_k\int\chi_k f\,\dd^2x$ is then exact, and each term can be integrated on a grid adapted to it.} Lens centres other than $\bs{c}$ (e.g.\ the second mass of a binary) are singular points that a polar grid about $\bs{c}$ cannot resolve. Around each of them we therefore integrate a patch on its own polar grid, weighted by $\chi_k = S(|\bs{x}-\bs{x}_k|/\rho_k)$, while the main grid carries $\chi_0 = 1-\sum_{k\ge1}\chi_k$. Here $S$ is the infinitely differentiable Planck-taper step \cite{McKechan:2010kp}, so the split introduces no artificial edges. The head is then 
\begin{equation}
\int_{|\bs{x}-\bs{c}|\le r_b} e^{iwT}\,\dd^2\bs{x}\;\approx\;e^{iw|\bs{y}|^2/2}\sum_j W_j\,e^{iw\,(A_j-\bs{x}_j\cdot\bs{y})}\,
\end{equation}
where the quantity $A_j$ is independent of $w$ and $y$ and has the form 
\begin{equation}
A_j = \frac{|\bs{x}_j|^2}{2}-\psi(\bs{x}_j)\,,
\end{equation}
where the sum runs over the nodes of the main grid and of all patches. The nodes and weights depend neither on $w$ nor on $\bs{y}$, so $\psi$ is evaluated only once for all nodes. For many frequencies the sum is a type-3 non-uniform discrete Fourier transform (sources $A_j-\bs{x}_j\cdot\bs{y}$, targets $w$), computed with \texttt{FINUFFT} \cite{Barnett2019_finufft}; for many source positions at one frequency it is a two-dimensional one.

On the vertical path $u = u_b + it$, write $T = U + iV$. The modulus of the integrand is
\begin{equation}\label{eq:decay_condition}
|e^{i w T(u_b + it)}| = e^{-w V(t)}\,,
\end{equation}
so the tail decays if $V$ grows with $t$. By the Cauchy--Riemann equations, the initial growth rate of $V$ along the vertical path equals the slope of $T$ along the real axis at the split point,
\begin{equation}
\alpha(\theta) \equiv \frac{\partial V}{\partial t}\Big|_{t=0} = \frac{\dd T}{\dd u}\Big|_{u_b} = 1 - \frac{\hat{\bs{n}}\cdot\bs{b} + \partial_r \psi(\bs{c} + r_b\hat{\bs{n}})}{r_b}\,.
\end{equation}
Setting $\dd T/\dd u = 0$ gives the radial component of the lens equation, so $\alpha>0$ means that no image lies beyond the split along that direction. We choose $r_b$ such that $\alpha \ge 1/2$ in every direction,
\begin{equation}
\max_\theta\, \frac{\hat{\bs{n}}\cdot\bs{b} + \partial_r \psi(\bs{c} + r_b\hat{\bs{n}})}{r_b} \leq \frac{1}{2}\,,
\end{equation}
which places all images inside the head and makes the tail decay at least as fast as $e^{-wt/2}$. To obtain the tail integral in Gauss--Laguerre form, consider $\Delta T(t) = T(u_b + it) - T(u_b)$, where $T(u_b)$ is real. The tail becomes
\begin{equation}\label{eq:tail_int_identity}
 \int_{u_b}^{\infty} e^{i w T(u)}\, \dd u = i\, e^{i w T(u_b)}\int_0^{\infty}e^{i w \Delta T(t)}\,\dd t\,.
\end{equation}
Since $\Delta T \approx i\alpha t$ near the split, the substitution $\xi = \alpha w t$ brings it to Gauss--Laguerre form,
\begin{equation}
\frac{i\, e^{i w T(u_b)} }{\alpha w }\int_0^{\infty} e^{-\xi}\, G(\xi)\,\dd \xi
\approx  \frac{i\, e^{i w T(u_b)}}{\alpha w} \sum_{j = 1}^{n} W^{\rm L}_j\, G(\xi_j)
\end{equation}
where the factor $G(\xi)$ is
\begin{equation}
    G(\xi) = e^{\,i w \Delta T(\xi/\alpha w) + \xi}\,
\end{equation}
The factor $e^{\xi}$ removes the leading decay, so $G$ contains only the slowly varying part of the integrand, which the rule integrates to high accuracy \cite{Huybrechs2015}. The nodes $\xi_j$ are the roots of the Laguerre polynomial $L_n$ and $W^{\rm L}_j$ the corresponding weights.\footnote{The default is \texttt{nlag = 30} Gauss--Laguerre nodes obtained from \texttt{scipy.special.roots\_laguerre}.} The tails of all directions are combined with the same trapezoid weights $2\pi/N$ as the head.

Because the nodes $t_j = \xi_j/(\alpha w)$ depend on $w$, $G$ is tabulated at a few Chebyshev values of $1/w$ in each frequency band and interpolated, so that $\psi$ on the complex path is again evaluated only once. Head and tail together give $F$ for all frequencies and source positions, with an error that depends only on the numbers of nodes.
\begin{figure}[ht]
    \centering
    \includegraphics[width=1.0\linewidth]{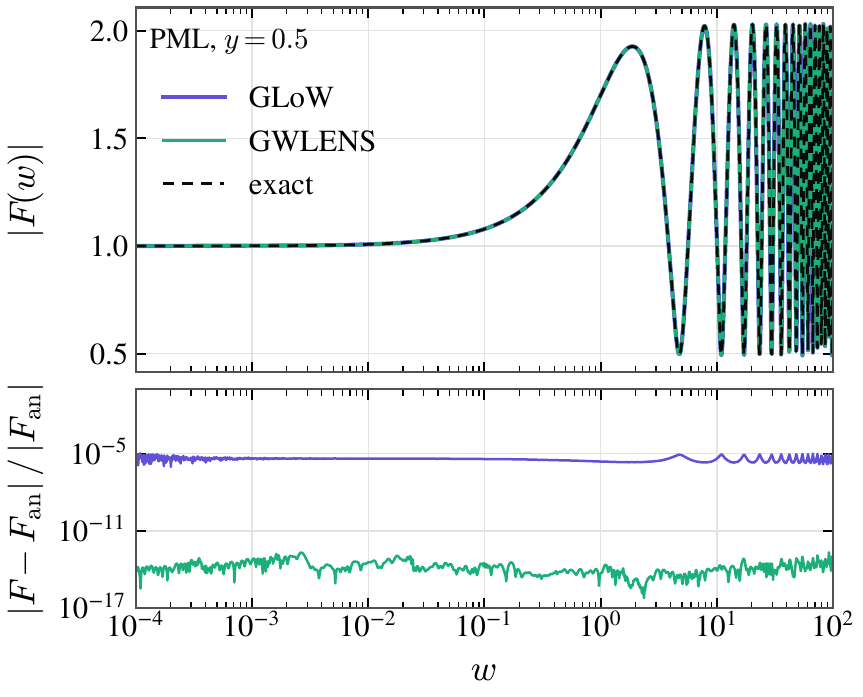}
    \caption{\textit{Top:} A comparison of exact PML amplification factor with GWLENS and GLoW for $y=0.5$. \textit{Bottom:} Relative error with respect to exact solution. }
    \label{fig:point-mass-accuracy-comparison}
\end{figure}

\begin{figure*}[ht]
    \centering
\includegraphics[width=1.0\linewidth]{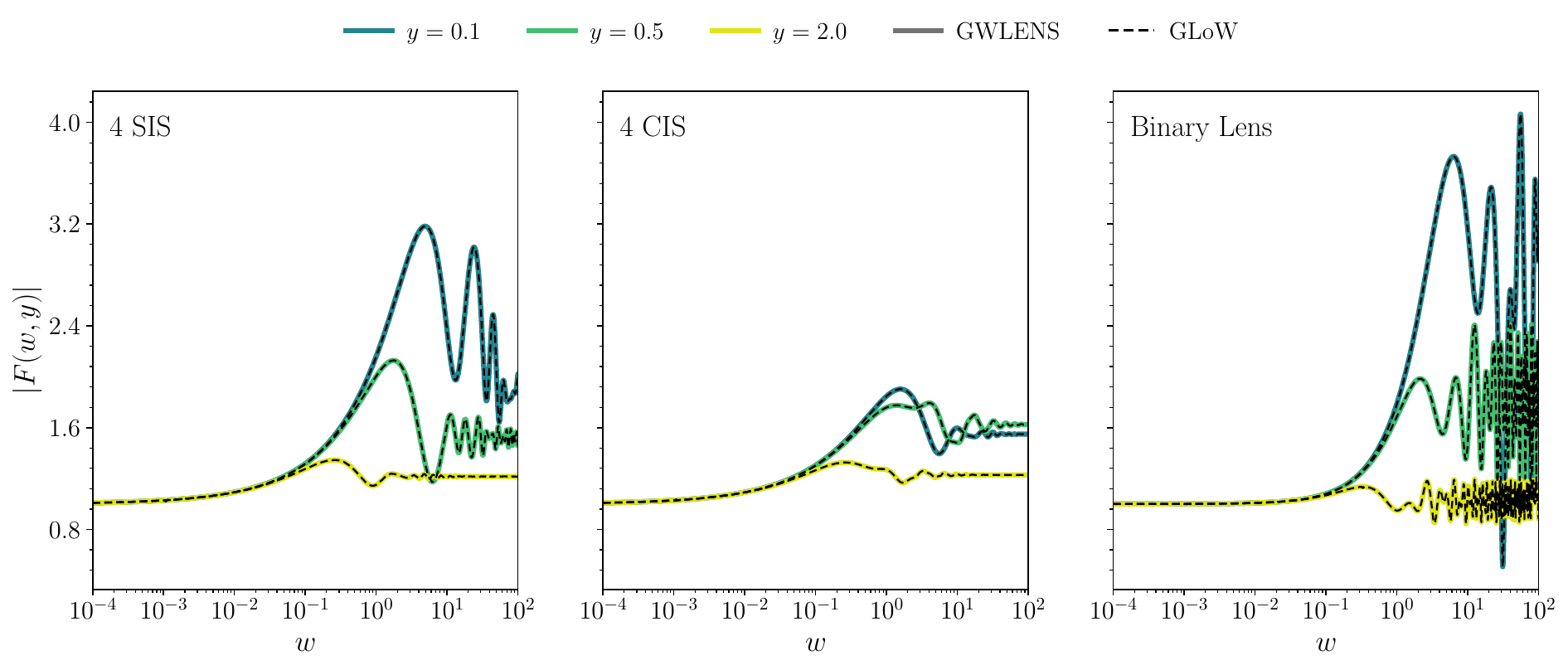}
    \caption{Amplification factor $|F(w,y)|$ for three different non-axisymmetric lens models: 4 SIS, 4 CIS and an equal mass binary point lens for different impact parameters. Solid curves represent \texttt{GWLENS}, while dashed black curves represent \texttt{GLoW} results. The SIS and CIS lenses are randomly distributed and the core radius of CIS lenses are chosen to be $x_c = 0.05$. The distance between two point masses in the binary are taken to be $d= 1.0$ along the x-axis placed at points $(-d/2,0)$ and $(d/2,0)$ respectively.}
    \label{fig:2dlens-comp}
\end{figure*}

\subsection{Axisymmetric lenses}\label{subsec:axially_sym}

\begin{figure*}[ht]
    \centering
\includegraphics[width=1.0\linewidth]{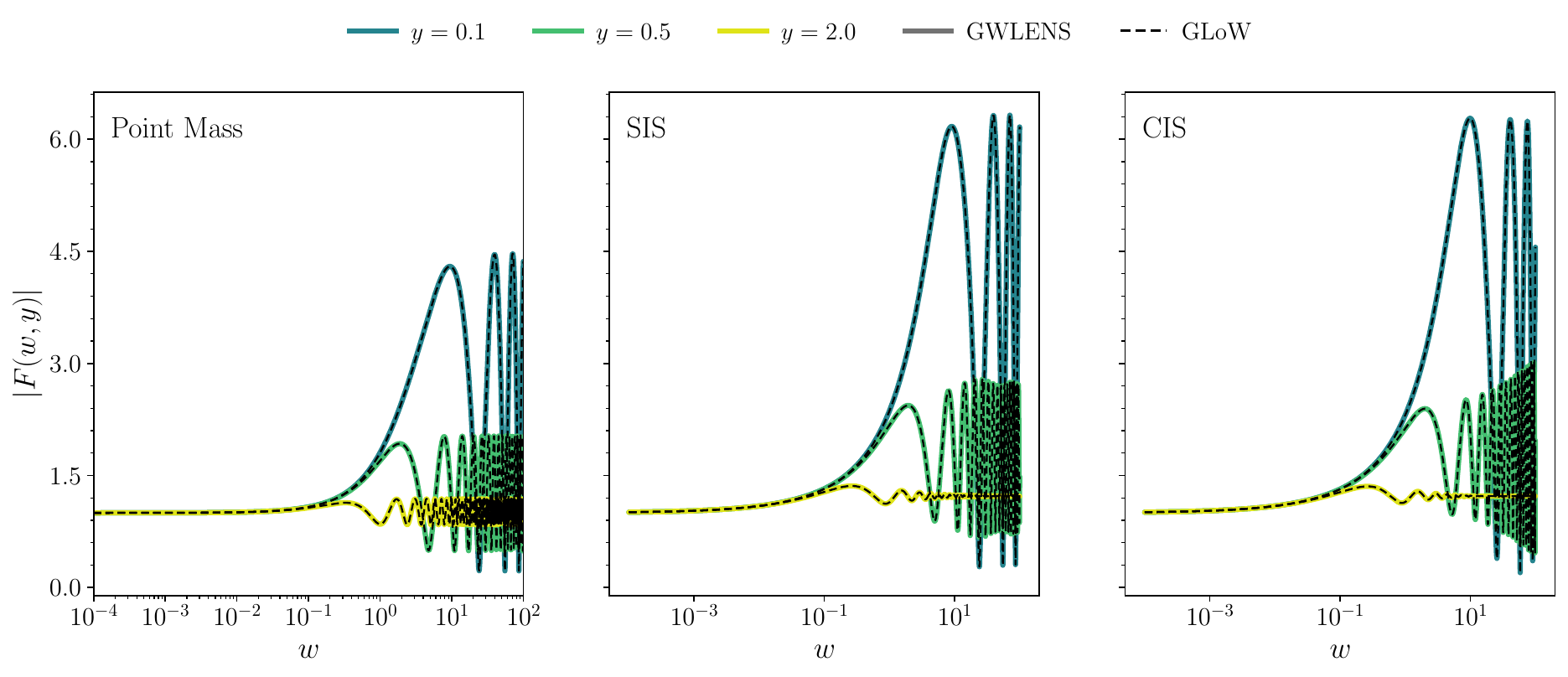}
    \caption{Amplification factor $|F(w,y)|$ for axisymmetric lens models: point mass, SIS and CIS for different impact parameters. Solid curves represent \texttt{GWLENS} and dashed lines are computed with \texttt{GLoW}. The core radius of $x_c= 0.05$ has been adopted for a CIS lens.}
    \label{fig:1dlens-comp}
\end{figure*}
For axially symmetric lens models, the lensing potential $\psi(\bs x)$  depends only on the radial distance $|\bs x|=r$ and we set the centre $\bs{c} = 0 $. The diffraction integral becomes
\begin{equation}\label{equation:amp_integral}
F(w,y) = -i w\, e^{i w y^2/2}\int_0^\infty r J_0(w r y)\exp\Big[i w\Big(\frac{r^2}{2} - \psi\Big)\Big]dr,
\end{equation}
where the $J_0$ is the zeroth-order Bessel function, obtained after angular integration. Substituting $u = r^2/2$, the amplification factor reads
\begin{equation}
F(w,y) = -i w\, e^{i w y^2/2} \int_0^\infty f(u)\, e^{i w u}\, du\,, 
\label{eq:F1d}
\end{equation}
where $f(u) = J_0\!\left(w y \sqrt{2u}\right) e^{-i w \psi(\sqrt{2u})}.$
The integrand does not decay at large $u$ and converges only
conditionally; truncating it at a finite cut-off introduces errors that
depend on both the cut-off and $w$. As detailed in Sec.~\ref{subsec:general_method}, we split
Eq.~(\ref{eq:F1d}) at $u_b = x_b^2/2$, chosen beyond all images such that the decay rate $\alpha \geq 1/2$ is satisfied. On the vertical path, $du = i dt$ and $e^{iwu} = e^{iwu_b}e^{-wt}$. Hence the amplification factor reads
\begin{equation}
\begin{split}
F(w,y) &=\ - iw\,e^{iwy^2/2}\Big[\underbrace{\int_0^{u_b} f(u)\,e^{iwu}\,du}_{\text{head: real axis}} \;+\; \\ 
&+\ \underbrace{i\,e^{iwu_b}\int_0^{\infty} f(u_b+it)\,e^{-wt}\,dt}_{\text{tail: complex plane}}\Big]\,.
\end{split}
\label{eq:Faxi}
\end{equation} 
On the tail, the growth of the Bessel function $J_0$ is outweighed by $e^{-wt}$. The integrand is analytic and decays in the upper quarter plane for $\mathrm{Re}\,u \ge u_b$. The finite head $[0, u_b]$ is now integrated with  composite Gauss--Legendre quadrature on sub-intervals containing a fixed set of phase cycles as in the general case. The maximum possible phase accumulated up to a point $u_b$ is given by
\begin{equation}
\phi(u) = \int_0^{u} \left|\frac{dA}{du'}\right| + y_{\mathrm{max}}\ \sqrt{2u}\,. 
\end{equation}
where $A(u) = u - \psi(\sqrt{2 u}) $ and the second term is the contribution from the Bessel function which oscillates like $e^{\pm i w y \sqrt{2u}}$. In the worst case we take the $+$ sign, i.e., $+y_{\mathrm{max}}\sqrt{2u}$. At a fixed $w$, the integrand has gone through $w \phi(u)/2\pi$ cycles and $[0,u_b]$ is divided into sub-intervals each containing a fixed number of these cycles. The head is now the weighted sum as in the general case
\begin{equation}
 \int_0^{u_b} f(u)e^{i w u}du \approx \sum_j W_j J_0(w y \sqrt{2 u_j}) e^{i w A_j}\,,
\end{equation}
The tail decays exponentially and is integrated with a scaled Gauss--Laguerre rule. This form needs only values of $\psi$ at complex points, and its error falls exponentially with the number of quadrature nodes. Unlike Picard--Lefschetz methods, no saddle points need to be located: only the part of the contour beyond the images is deformed, and the head stays on the real axis. As a sanity check we compare the accuracy of our method with the Point Mass Lens (PML) analytical solution \cite{Takahashi2003} with the numerical solution from {\gwlens} and {\glow} in Fig.~\ref{fig:point-mass-accuracy-comparison}. {\gwlens}  agrees with the analytical solution to a relative error of ${\approx}10^{-13}$, several orders of magnitude below {\glow} at the settings used (\texttt{tmin: 1e-6, tmax: 1e10, Nt: 50000 }). 

\section{Code structure}\label{sec:code-structure}

\begin{figure*}[ht]
    \centering
    \includegraphics[width=0.85\linewidth]{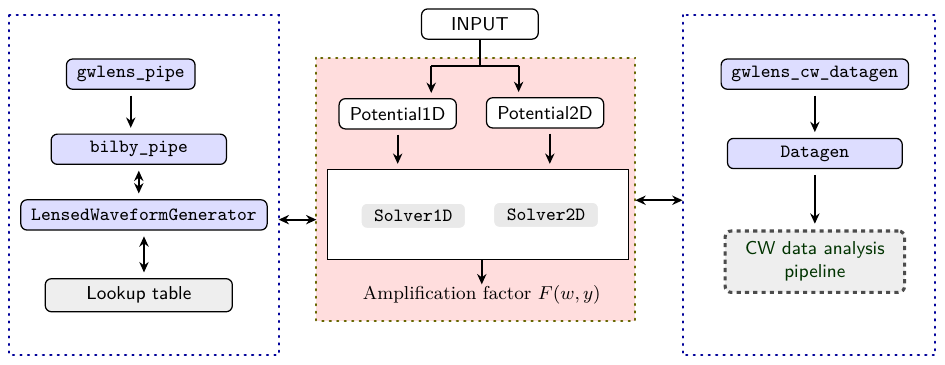}
    \caption{Code structure of {\gwlens} divided into flow diagram blocks: a main diffraction integral solver (middle), communicating with the CBC parameter estimation (left), and the CW data generation (right), coupled with a CW data analysis pipeline (e.g. the  Time-Domain $\mathcal{F}$-statistic method, \texttt{TDFstat}).}
    \label{fig:gwlens_code_structure}
\end{figure*}

{\gwlens} consists of three main components: a lensing diffraction integral solver based on the method described in Sec.~\ref{sec:twisted-tail}, a command-line tool for PE of short-duration signals, and a component to generate time-domain data segments for CW lensing studies as illustrated in Fig.~\ref{fig:gwlens_code_structure}. The solver consists of {\solver} and {\solverb} which solve axisymmetric lensing potentials and general lensing configurations respectively. The \texttt{Potential1D} and \texttt{Potential2D} classes in these solvers allow users to write and input their arbitrary lens configuration and check for analyticity as the method demands the functions to be analytic in the tail region. The \texttt{solver1d.amplification(w, y, lens, **solver\_kwargs)} and the analogous function in {\solverb} generate the lensing amplification factor.

The second component of our pipeline is \texttt{gwlens\_pipe}, an extension of \texttt{bilby\_pipe} that passes the lensing parameters to the \texttt{LensedWaveformGenerator} class in {\gwlens}. This allows users to generate, inject and recover lensed signals with \texttt{Bilby} \cite{bilby_paper, bilby_doi}. The waveform generator also supports an optional lookup table for faster parameter estimation of expensive lens models. By default, direct use of the solver for PE uses a reduced set of quadrature nodes\footnote{Specifically, the PE settings are: \texttt{ngl:24, nlag:30}.} for a faster evaluation compared to default settings. It has been verified that these settings result in a relative error of $10^{-8}$, compared to exact analytic functions. 

The third component of {\gwlens} is aimed at studies relating to lensing of long-duration sources, in particular the CW sources, e.g. rotating, non-axisymmetric neutron stars (NSs). The command-line tool \texttt{gwlens\_cw\_datagen} reads an \texttt{*.ini} file with the parameters needed to generate data for further data analysis tasks, e.g. the all-sky Time-Domain $\mathcal{F}$-statistic  (\texttt{TDFstat}) search pipeline \cite{JKS98,Astone2010,tdfstat,Aasi_2014,PhysRevD.96.062002,PhysRevD.97.102003,PhysRevD.100.024004,PhysRevD.106.102008}. For more details we refer the reader to Appendix~\ref{app:cw_lensing} and the references therein. 

\section{Performance and Accuracy}\label{sec:benchmarking}

\begin{figure}[ht]
    \centering
    \includegraphics[width=1.0\linewidth]{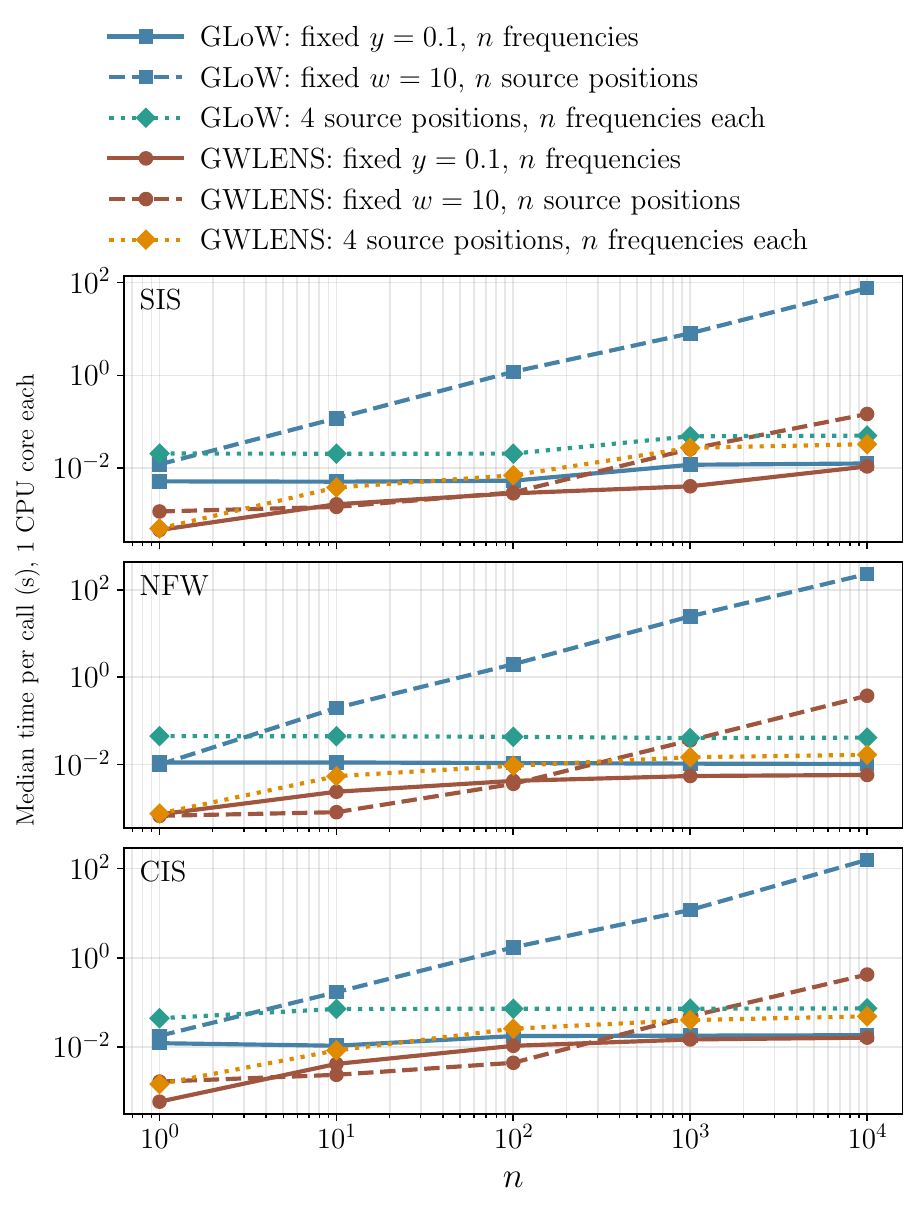}
     \caption{ A single CPU core runtime for {\gwlens}  (\texttt{solver1d}) and \texttt{GLoW} (\texttt{SingleIntegral}) for symmetric lenses SIS, NFW ($x_s = 0.1$) and a CIS ($x_c = 0.05$). Solid lines  show the cost versus the number of frequencies $n=n_w$ in the range $[10^{-4},10^2]$ at  fixed $y= 0.1$; dashed lines show the median time \textit{vs} the number of source positions $n= n_y$ in the range $[0.1,1.0]$ at a fixed  $w=10$. The dotted lines show evaluation where $n=n_w$ frequencies are evaluated for four different source positions per call.}
    \label{fig:benchmark1d_1cpu}
\end{figure}

\begin{figure}[ht]
    \centering
    \includegraphics[width=1.0\linewidth]{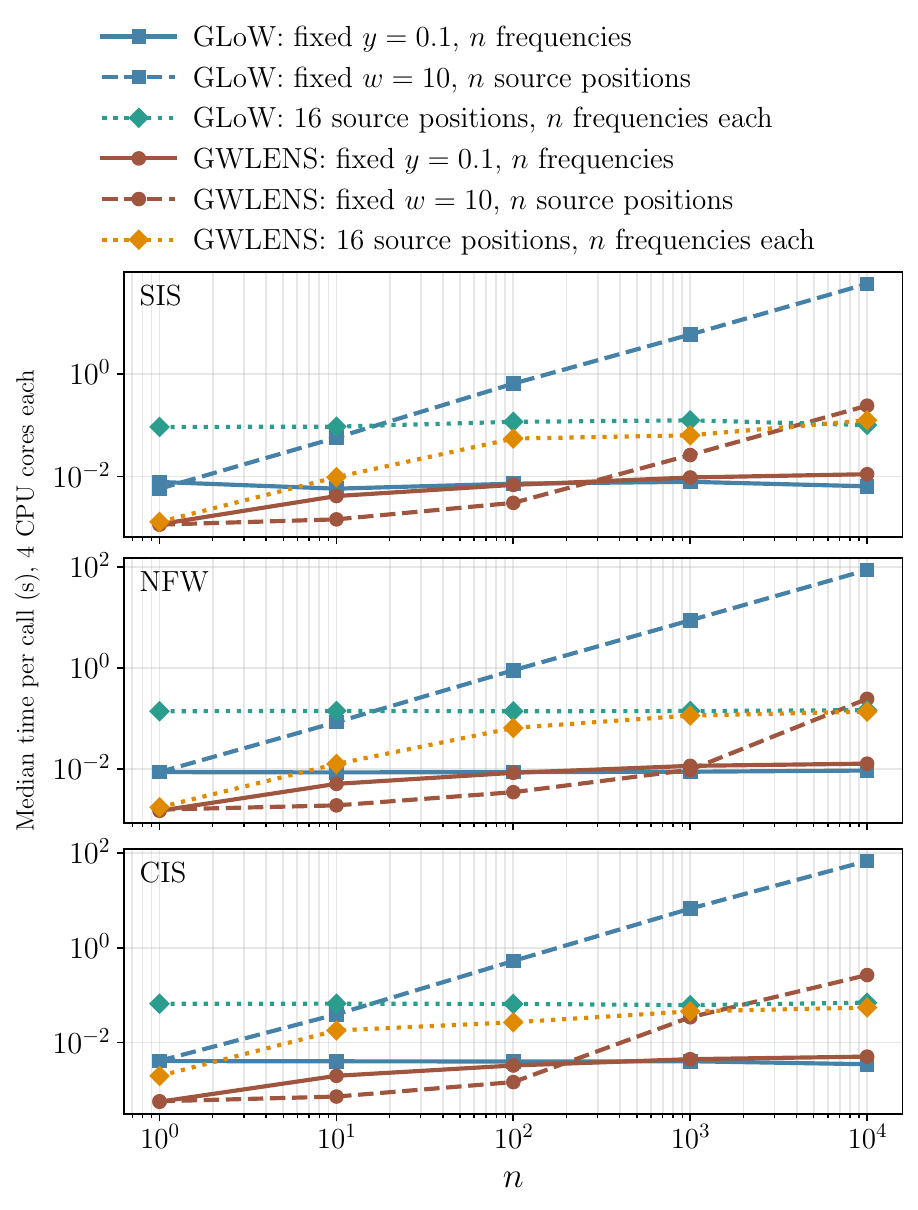}
     \caption{ A $4$ CPU core runtime for {\gwlens}  (\texttt{solver1d}) and \texttt{GLoW} (\texttt{SingleIntegral}) for symmetric lenses SIS, NFW ($x_s = 0.1$) and a CIS ($x_c = 0.05$). Solid lines show the cost versus the number of frequencies $n=n_w$ in the range $[10^{-4},10^2]$ at  fixed $y= 0.1$; dashed lines show the median time \textit{vs} the number of source positions $n=n_y$ in the range $[0.1,1.0]$ at a fixed $w=10$. The dotted lines show evaluation where $n=n_w$ frequencies are evaluated for 16 different source positions per call.}
    \label{fig:benchmark_single}
\end{figure}

\begin{figure}[ht]
    \centering
    \includegraphics[width=1.0\linewidth]{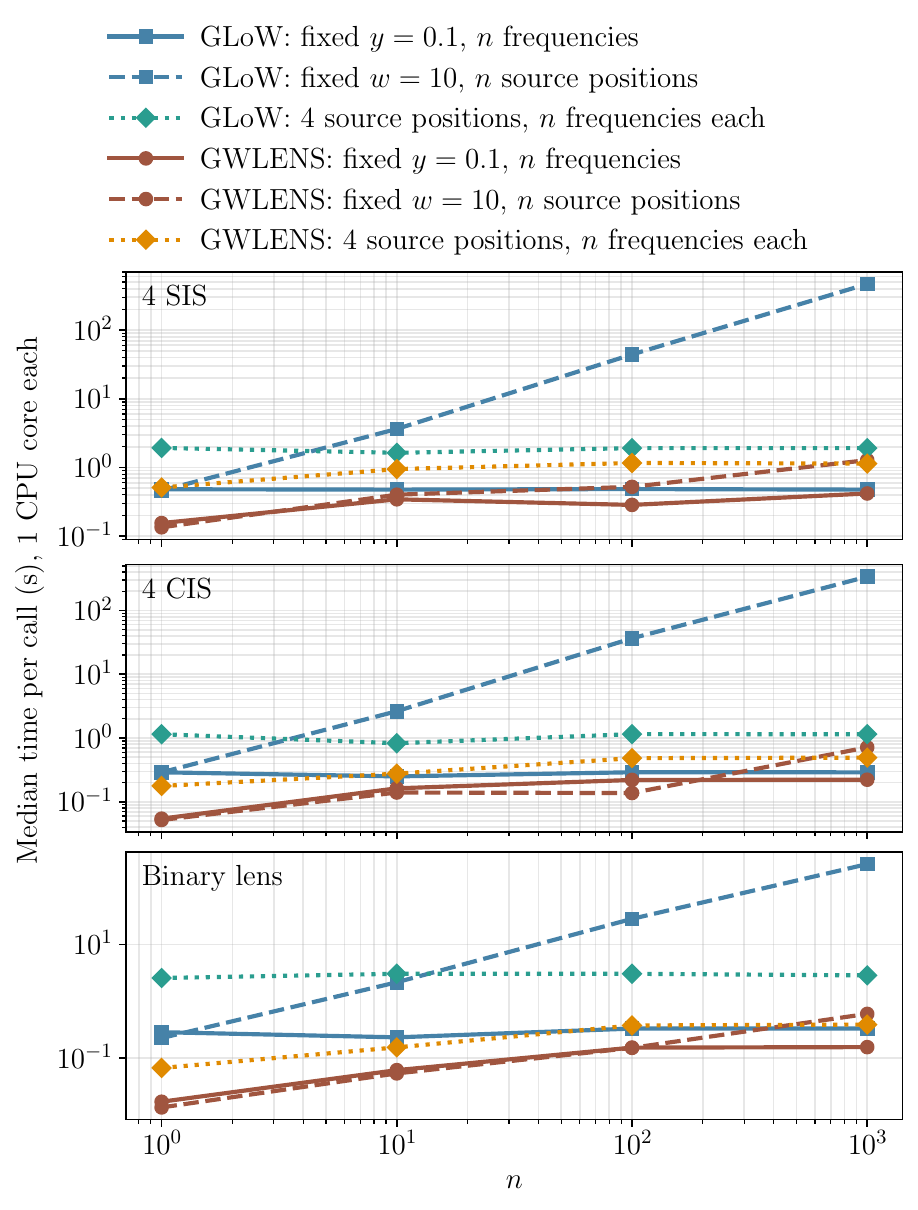}
     \caption{A single CPU core runtime for {\gwlens} (\texttt{solver2d}) and \texttt{GLoW} (\texttt{Multicontour}) non-axisymmetric lenses consisting of 4 SIS, 4 CIS and a binary lens. Solid lines  show the cost versus the number of frequencies $n=n_w$ in the range $[10^{-4},10^2]$ at  fixed $y= 0.1$; dashed lines show the median time \textit{vs} the number of source positions $n=n_y$ in the range $[0.1,1.0]$ at a fixed $w=10$. The dotted lines show evaluation where $n=n_w$ frequencies are evaluated for four different source positions per call.}
    \label{fig:benchmark2d_1cpu}
\end{figure}

\begin{figure}[ht]
    \centering
    \includegraphics[width=1.0\linewidth]{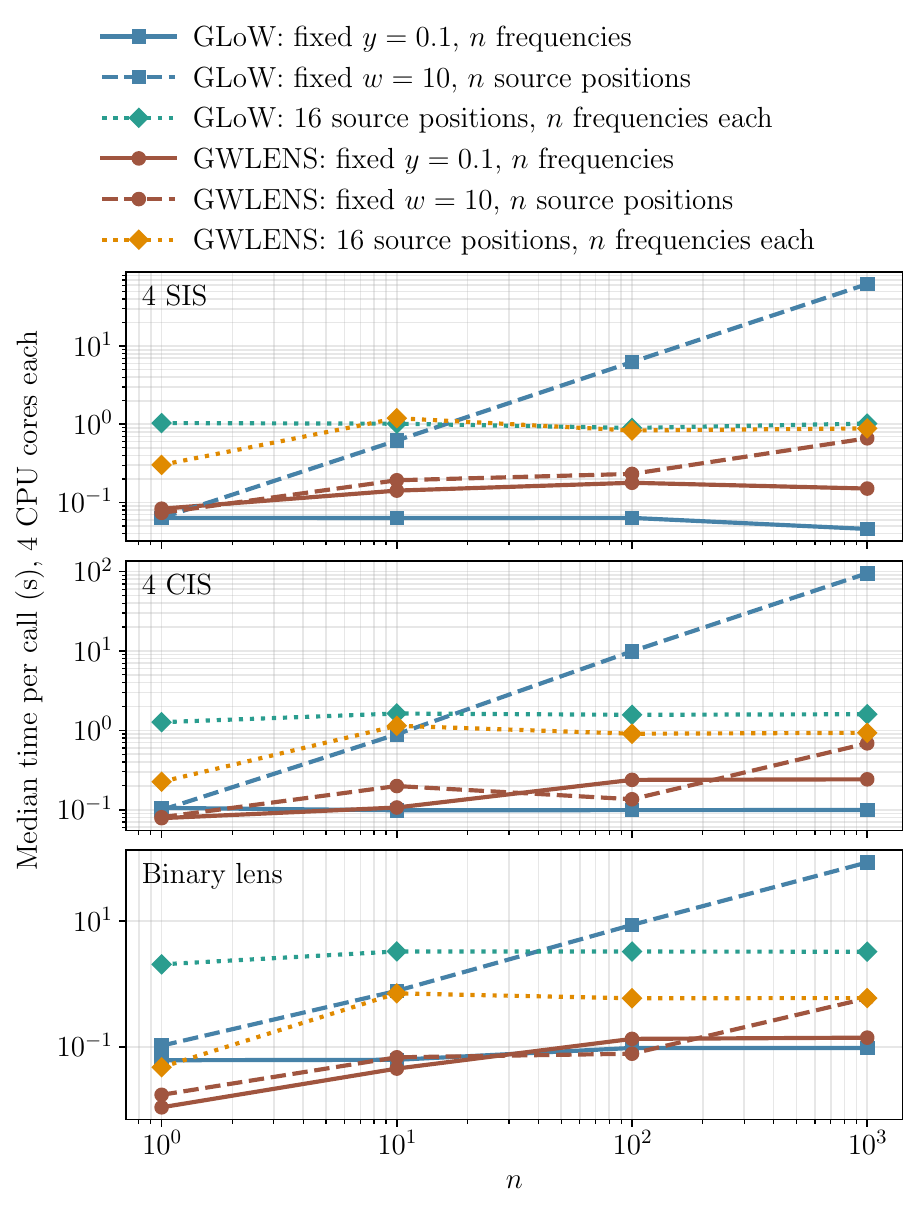}
     \caption{A $4$ CPU core runtime for {\gwlens} (\texttt{solver2d}) and \texttt{GLoW} (\texttt{Multicontour}) for non-axisymmetric lenses consisting of 4 SIS, 4 CIS and a binary lens. Solid lines  show the cost versus the number of frequencies $n=n_w$ in the range $[10^{-4},10^2]$ at  fixed $y= 0.1$; dashed lines show the median time \textit{vs} the number of source positions $n=n_y$ in the range $[0.1,1.0]$ at a fixed $w=10$. The dotted lines show evaluation where $n=n_w$ frequencies are evaluated for 16 different source positions per call.}
    \label{fig:benchmark2d_4cpu}
\end{figure}

In this section we compare the performance of the twisted-tail method implementation in {\gwlens}  with the contour integration in {\glow} for different lens models. For benchmarking we consider several 1D and 2D lens models with the same set of parameters and perform tests on the same hardware. The calculations were performed on a Linux-based desktop equipped with an AMD Ryzen 9 9950X processor, with 16 physical CPU cores and 32 hardware threads. We restrict the core count using \texttt{OMP\_NUM\_THREADS} and report the median time of five repeated evaluations of the same call and use the following precision settings in {\glow} (\texttt{tmin:1e-2, tmax:1e9, Nt:5000}) and for {\gwlens} we used (\texttt{ngl:24, nlag:24, n\_cheb:13, nufft\_eps:1e-9}).

For axially symmetric lens models we considered the following lens profiles: a Singular Isothermal Sphere (SIS), Cored Isothermal Sphere (CIS)  with a core radius fixed to $x_c = 0.05$ and a Navarro--Frenk--White (NFW) with $x_s = 0.1 $ (see Table~\ref{tab:lens_summary} for all profiles). The corresponding amplification curves for these lens models were shown in Fig. \ref{fig:1dlens-comp}. In {\gwlens} we evaluate them using \texttt{amplification} function in {\solver} and in {\glow} with {\glowsi}, which \cite{Villarrubia-Rojo:2024xcj} reports to be faster than {\glowmc} for axisymmetric lenses. As can be seen from Fig.~\ref{fig:benchmark1d_1cpu} {\gwlens} out-performs {\glow} in computing amplification factor for a fixed $w$ and many source positions by several orders. For a fixed $y$ and varying number of dimensionless frequencies $n = n_w$ both codes report a median run time less than $10^{-2}\,\mathrm{s}$ with {\gwlens} performing better than {\glow} for single core evaluations, and {\glow} performing better for multiple CPUs at large $n=n_w$; see Fig. \ref{fig:benchmark_single} for corresponding benchmarks. Parallelization of multiple contour integral evaluations is enabled by default in {\glow} with all available computing resources. On the other hand the parallelization in {\gwlens} is efficient in multiple source positions rather than multiple frequencies evaluation by construction, therefore {\gwlens} under-utilizes the available resources. To demonstrate this, we benchmark the performance for a single function call that evaluates multiple frequencies and 4 and 16 source positions for $1$ CPU and $4$ CPUs respectively. These evaluations are now distributed efficiently across available cores and the computing time is reported with dotted lines.

As for symmetric lenses, the general lenses  were benchmarked for a single (Fig. \ref{fig:benchmark2d_1cpu}) and $4$ CPU core (Fig. \ref{fig:benchmark2d_4cpu}) to evaluate their performance. We considered composite lens models formed with 4 SIS  and 4 CIS ($x_c = 0.05$) lenses randomly distributed and a binary point lens \cite{Serra:2025kbw} with a fixed separation $d = 1.0 $ and their components placed at positions $(-d/2, 0)$ and $ (+d/2, 0)$ in the lens plane. The amplification curves for these models are compared in Fig. \ref{fig:2dlens-comp}. In {\gwlens} we evaluate them using \texttt{amplification} function in {\solverb} and in {\glow} with {\glowmc}. As in symmetric lenses, {\gwlens} performs better for many source positions and fixed frequencies and shows comparable or better performance for a single core across different lens models considered. 

\begin{figure*}[ht]
    \centering    
    \includegraphics[width=0.90\textwidth]{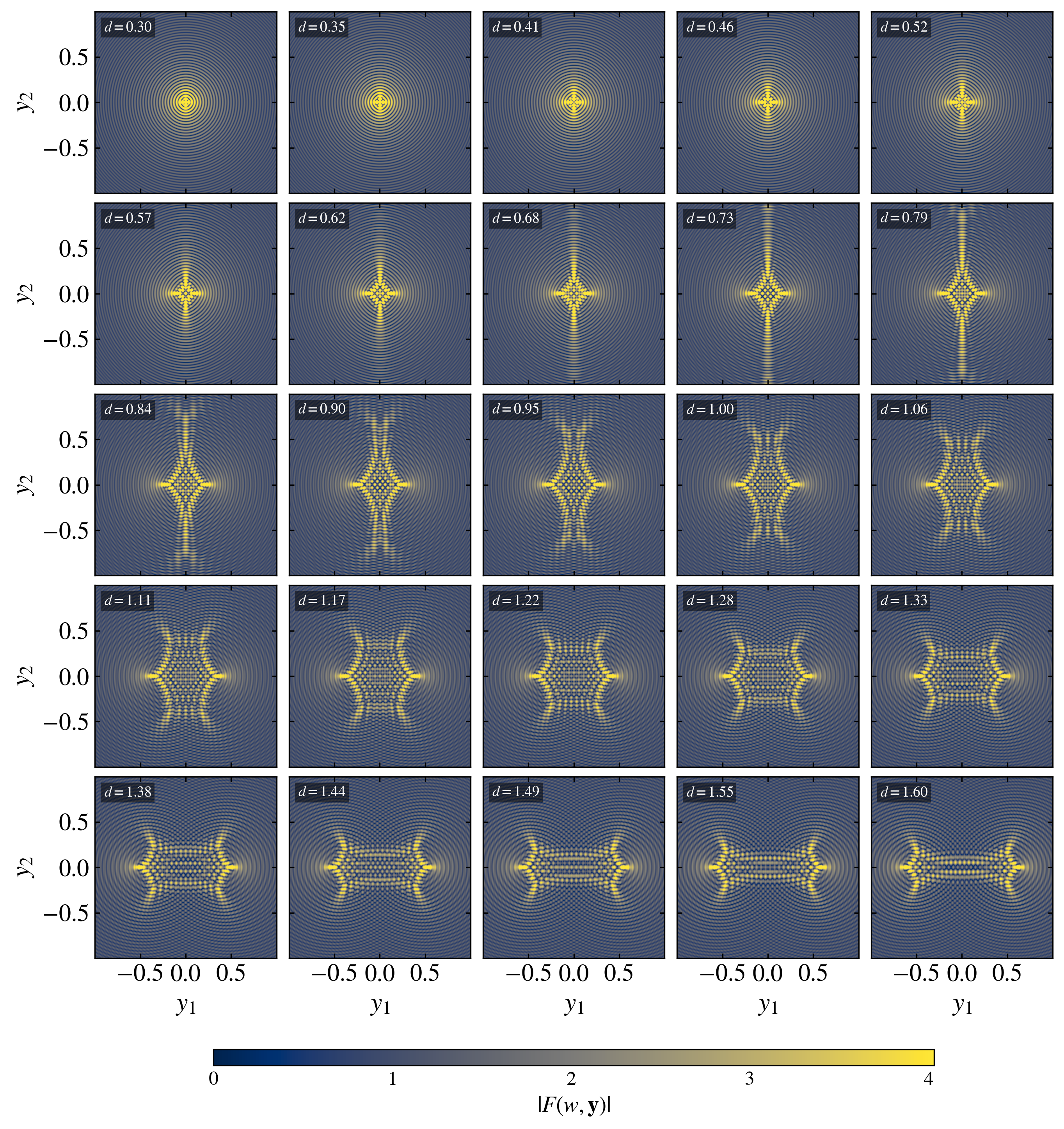}
    \caption{Amplification maps $|F(w,\boldsymbol{y})|$ of an equal-mass binary point lens (masses $0.5+0.5$) at $w=100$ for 25 separations $d\in[0.30,\,1.60]$, increasing from left to right and top to bottom. Each panel covers the source-plane square $-1.0\le y_1\le 1.0$ and $-1.0\le y_2\le 1.0$ with $500\times500$ pixels, and the lenses lie on the $x_1$ axis at $\pm d/2$. As $d$ grows, the compact cross-shaped structure opens into two separated cusped structures joined by a bridge, with wave-optics interference fringes throughout. All 25 maps ($6.25\times10^6$ pixels) were computed in $\sim8.3$\,s on a single core, including the construction of a new grid for every separation (using the lighter PE settings).}
    \label{fig:binary-map-25}
\end{figure*}

\begin{figure*}[h]
  \centering
  \includegraphics[width=0.90\textwidth]{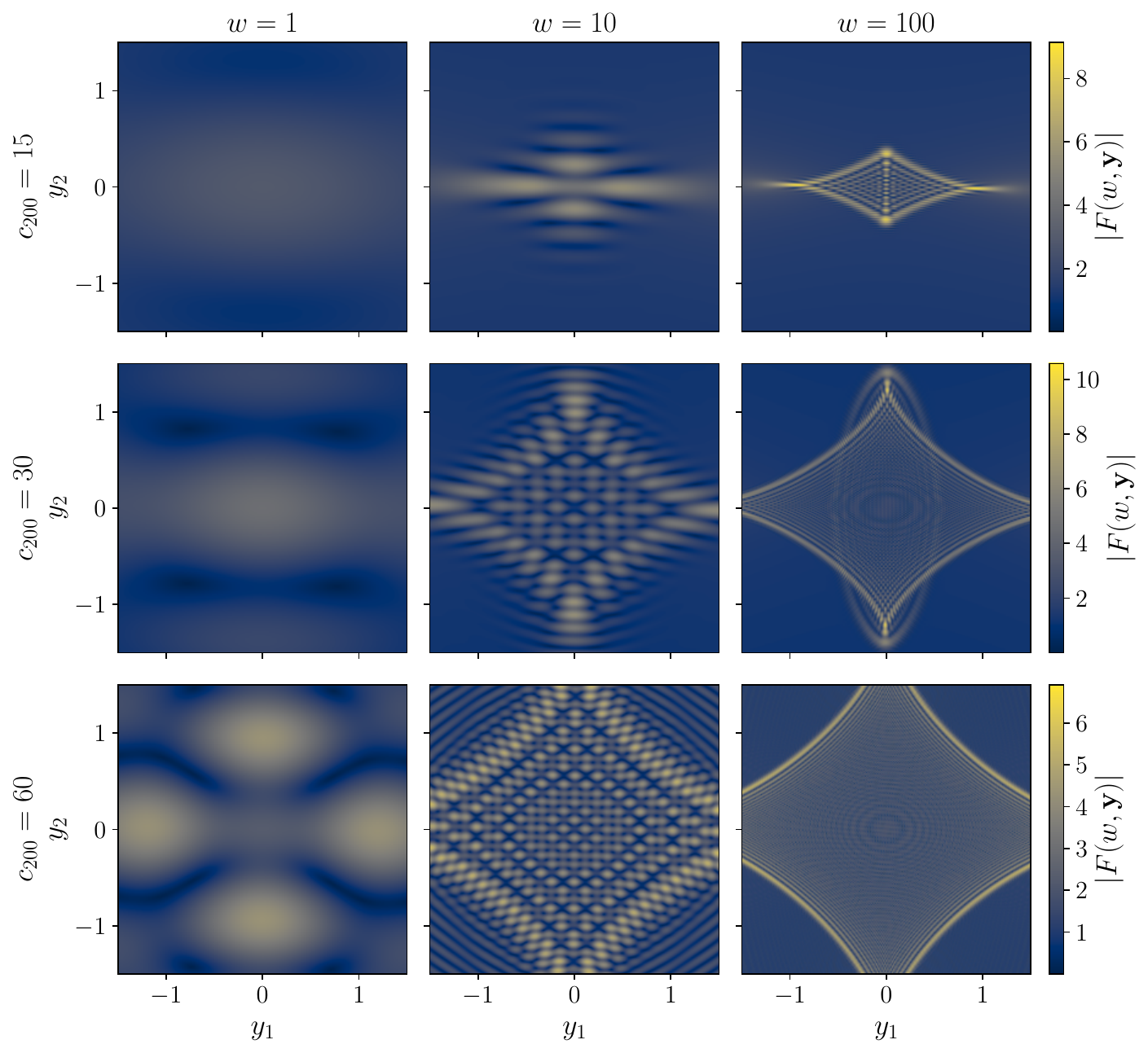}
  \caption{Amplification maps $|F(w,\boldsymbol{y})|$ of an elliptical NFW halo ($M_{200}=10^{13}\,M_\odot$, $z_l=0.5$, $z_s=1.5$, ellipticity $e_1=0.2$, $e_2=0$) in a weak external shear $(\gamma_1,\gamma_2)=(0.03,0.01)$, for concentrations $c_{200}=15$, 30, 60 (rows) and frequencies $w=1$, 10, 100 (columns). Each panel covers the source-plane square $-1.5\le y_1\le 1.5$ and $-1.5\le y_2\le 1.5$ with $500\times500$ pixels, and each row has its own colour scale. As $w$ increases the maps develop interference fringes whose extent grows with $c_{200}$. All nine maps ($2.25\times10^6$ pixels) took $\sim6.2$\,s on a single core, including the construction of a separate grid for every lens and frequency (using the lighter settings for parameter estimation). \texttt{FIONA} \cite{Ephremidze:2026era} reports ${\sim}130$\,s for the same lens configuration on a single CPU core (different hardware, so this result is indicative only).}
  \label{fig:enfw_maps}
\end{figure*}

\begin{figure*}[h]
    \centering
  \includegraphics[width=0.9\textwidth]{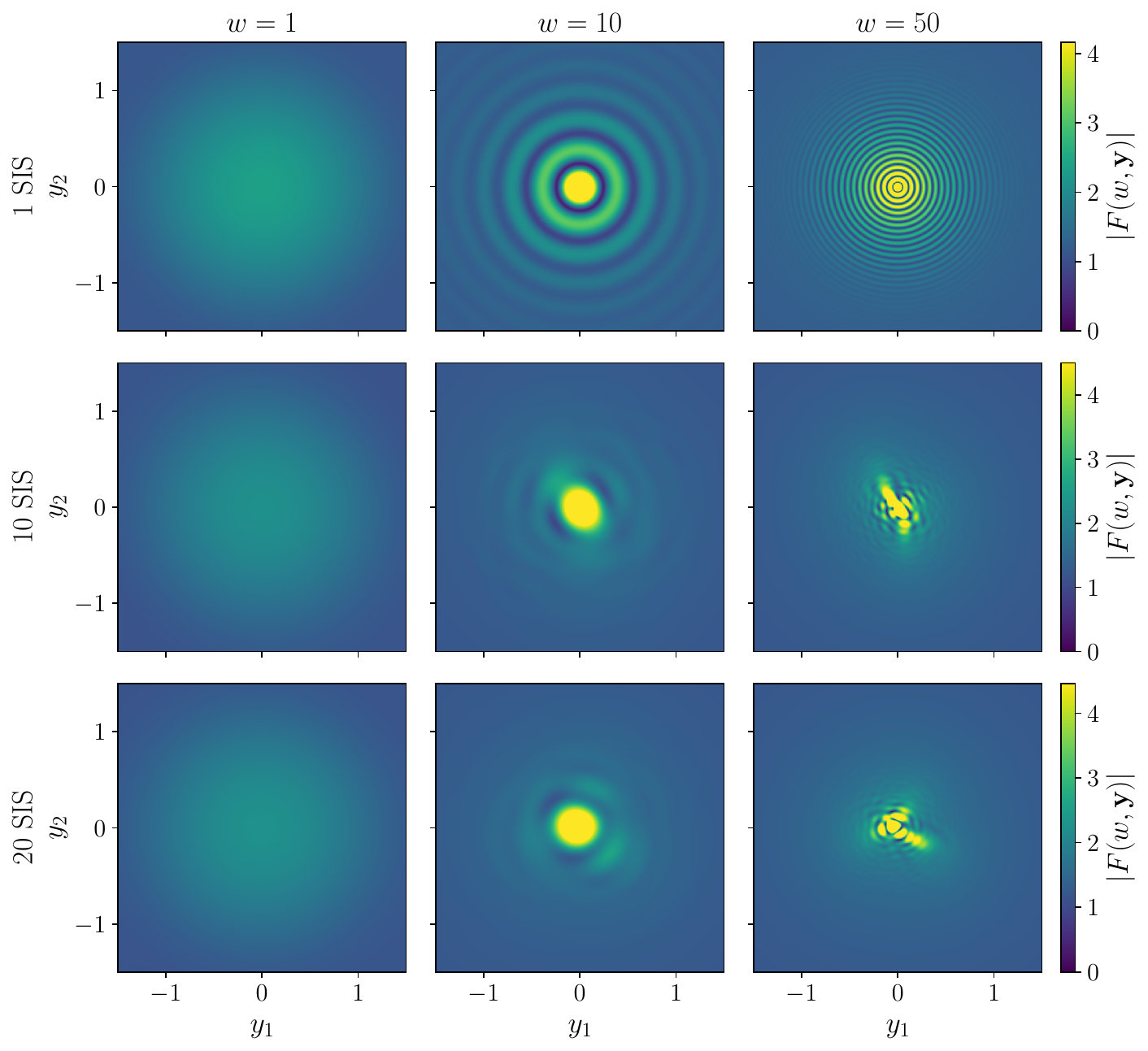}
  \caption{Amplification maps $|F(w,\boldsymbol{y})|$ of $N=1$, 10 and 20 singular isothermal spheres (rows) at $w=1$, 10 and 50 (columns). Each SIS has $\psi_0=1/N$, so the total $\psi_0$ is the same in all rows. The centres are drawn uniformly in a disc of radius 1 (fixed random seed) and shifted so that their centre of mass lies at the origin. Each panel covers the source-plane square $-1.5\le y_1\le 1.5$ and $-1.5\le y_2\le 1.5$ with $500\times500$ pixels, and each row shares one colour scale. For a single SIS the maps are concentric rings from the interference of the two images. For many SIS, the structure is confined to a small region near the origin, with irregular fringes from the several images and cusps. The nine maps ($2.25\times10^6$ pixels) took ${\sim}17$\,s on a single CPU core, using the lighter PE settings.}
  \label{fig:multi_sis}
\end{figure*}

The \texttt{amplification\_map} in {\solverb} is efficient in evaluating diffraction integral at many source positions. As an example, we calculate $25$ equal-mass binary point-mass maps with each map built on a $500 \times 500$ grid of positions as shown in Fig.~\ref{fig:binary-map-25} which took a total time of $8.3\,\mathrm{s}$. Each map is evaluated with a fixed dimensionless frequency $w = 100$ and varying separation between binary components. Similarly, for an elliptical NFW halo ($M_{200}=10^{13}\,M_\odot$, $z_l=0.5$, $z_s=1.5$, ellipticity $e_1=0.2$, $e_2=0$) in a weak external shear $(\gamma_1,\gamma_2)=(0.03, 0.01)$ is shown in Fig. \ref{fig:enfw_maps} which took a median time of ${\approx}6.2\,s$  compared to $~130\,s$ reported in \texttt{FIONA} \cite{Ephremidze:2026era}. In addition, we evaluate a multi-component SIS (Fig.~\ref{fig:multi_sis}) with $1$, $10$ and $20$ uniformly distributed lenses; the nine maps took $~17\ \mathrm{s}$ (median).

\section{Applications}\label{sec:applications}
We demonstrate the application of {\gwlens} to two classes of GW signals: CBCs and CWs. These applications illustrate the use of {\gwlens} for both PE of transient signals and the generation and analysis of lensed long-duration signals. For CBC signals, we present an integration with the \texttt{bilby\_pipe} \cite{bilby_pipe_paper} that enables lensed signal parameter estimation. For CW signals, we describe the generation of lensed data incorporating the time-dependent amplification induced by microlensing and demonstrate how the resulting signal modulation can be recovered using standard CW search methods.

\subsection{CBC Parameter Estimation} \label{sec:PE}

For CBC GW signals, {\gwlens} provides \texttt{gwlens\_pipe}, a wrapper around \texttt{bilby\_pipe} \cite{bilby_pipe_paper}. As shown in Fig.~\ref{fig:gwlens_code_structure}, \texttt{gwlens\_pipe} calls the lensing amplification factor modules from the core {\gwlens} package and provides a \texttt{LensedWaveformGenerator}. To keep it user-friendly, \texttt{gwlens\_pipe} uses the same \texttt{*.ini} configuration file format as \texttt{bilby\_pipe}, with a lensing section in which the user inputs lens specific arguments. This serves as a convenient tool to perform lensed signal model selection and PE. The lensing section of the configuration file contains the following settings:

\begin{itemize}[itemsep=1pt, topsep=1pt, parsep=1.5pt]
    \item \texttt{lens-model}: Lensing models built into {\gwlens} or user-defined lens models.
    \item \texttt{lens-solver}: Specifies whether a 1D or 2D solver is used.
    \item \texttt{lensed-injection}, \texttt{lensed-recovery}: Booleans that control whether lensing is applied to the injection, recovery, or both.
    \item \texttt{use-lookup-table}: Boolean that generates a precomputed lookup table for faster PE.
\end{itemize}

The lensing parameters must be added to the injection dictionary of the \texttt{*.ini} file, and the corresponding priors must be added to the prior file, in the same way as for the other source parameters.


\begin{figure*}[ht]
    \centering
    \includegraphics[width=0.9\textwidth]{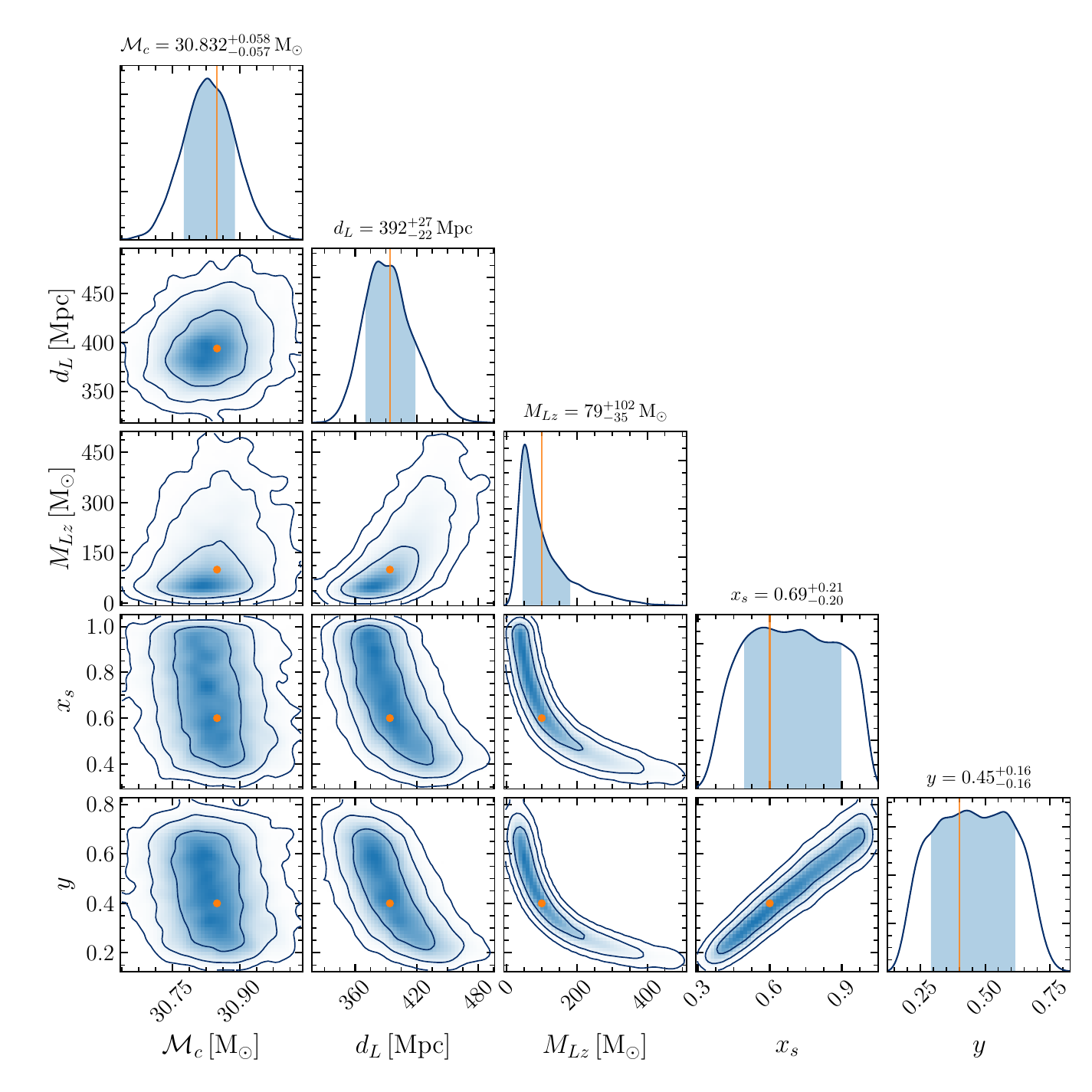}
    \caption{Posterior distributions from the \texttt{gwlens\_pipe} implementation for a GW250114-like aligned-spin signal, microlensed by an NFW lens, with microlensing parameters $(M_{Lz},\, y, x_s) = (100\, M_{\odot}, \, 0.4, 0.6)$. The parameter $d_L$ is the luminosity distance of the injected GW signal (see \cite{gw250114} for more details). The two-dimensional panels show the $1\sigma$, $2\sigma$ and 3$\sigma$ contour lines, and the histogram shows the $1\sigma$ error. The orange colour indicates the injection parameters.}
    \label{fig:gw250114-nfw}
\end{figure*}

As an example of \texttt{gwlens\_pipe}, we perform PE on a software injection of a GW250114\_082203-like signal (henceforth GW250114-like) \cite{gw250114} with aligned spins, with both lensed injection and recovery enabled. We choose an NFW halo as the lens with $M_{Lz} = 100\; M_{\odot}$ with $x_s  = 0.6$ and $y = 0.4$. The unlensed signal was modelled with the IMRPhenomXPHM approximant~\cite{Pratten:2020ceb} with aligned spins, and injected into Gaussian noise coloured by the projected
O4 (``O4high'') power spectral densities\footnote{O4 projected  sensitivity noise curves: \url{https://dcc.ligo.org/public/0165/T2000012}} of a three-detector network comprising LIGO Hanford (H), LIGO Livingston (L) and Virgo (V)~\cite{ALIGO2015, AdV2015}, as used in the observing-scenarios study~\cite{2020LRR....23....3A}. The signal is then recovered with the lensed waveform of the same lens model. We show the posteriors in Fig.~\ref{fig:gw250114-nfw}. For this PE study, we adopt the priors used for GW250114 in \cite{gw250114}, but restrict the spins to be aligned. The lensing priors are log-uniform in $M_{Lz} \in [10, 500]\, M_{\odot}$ and $x_s \in [0.2, 1]$, and a power-law prior on $y \in [0.05, 1]$ with an exponent of 1.
\begin{figure*}[ht]
    \centering
    \includegraphics[width=0.9\linewidth]{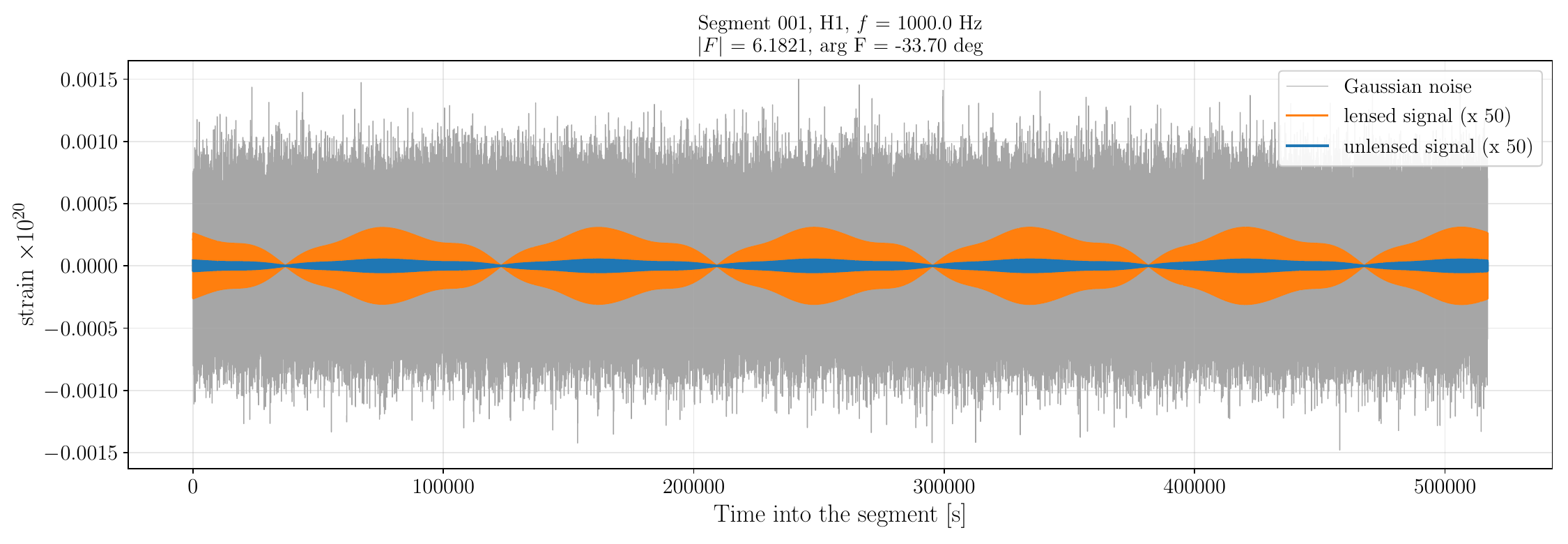}
    \includegraphics[width=1.0\linewidth]{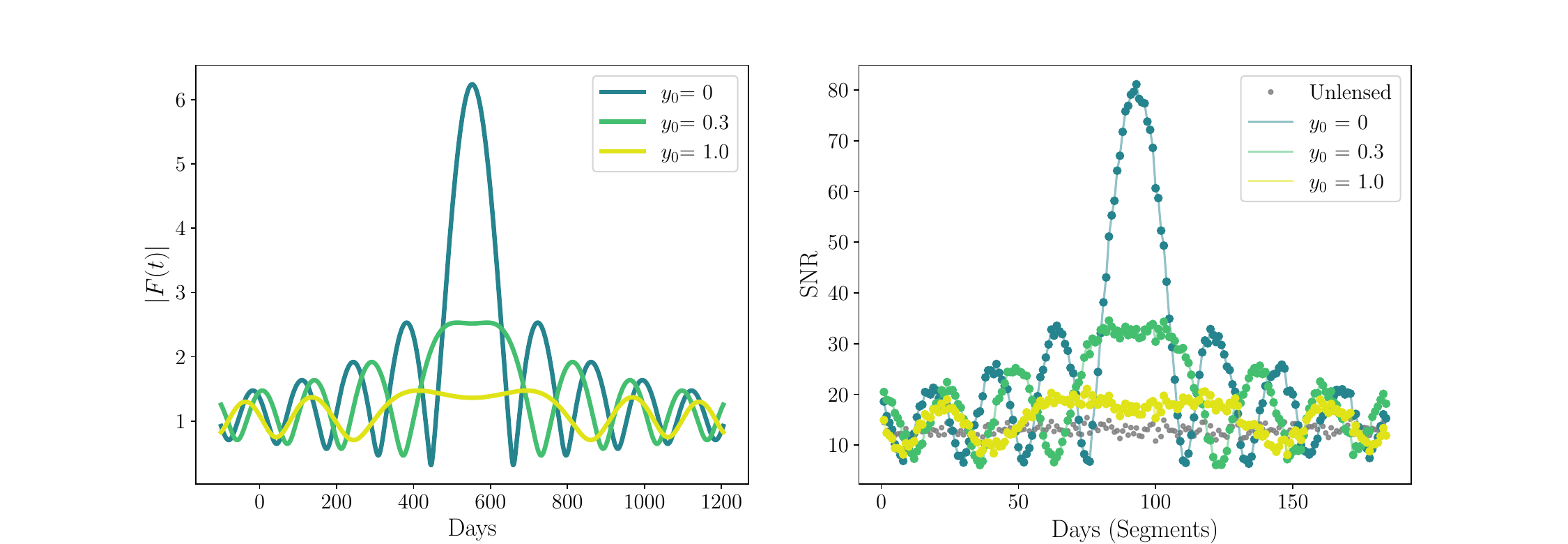}
    \caption{\textit{Top}: a six-sidereal-day segment (516985 samples)
generated with \texttt{gwlens\_cw\_datagen} around the peak of
the amplification, with bandwidth $B=0.5$\,Hz and reference
(lower band-edge) frequency $f=1000$\,Hz: a lensed CW signal
added to Gaussian noise. In this segment $F=6.18$ and
$\arg F=-33.70^\circ$. \textit{Bottom left}: $F(t)$ for a
$100\,M_\odot$ point lens at $D_L=10$\,kpc and a source at
$D_S=20$\,kpc, with relative velocity
$200$\,km\,s$^{-1}$, for three values of $y_0$.
\textit{Bottom right}: the corresponding time-domain
$\mathcal{F}$-statistic SNR, $\rho=\sqrt{2\mathcal{F}-4}$, in
each segment, evaluated at the injected signal parameters.}
    \label{fig:cwdata_pml_recovery}
\end{figure*}
\subsection{CW lensing}

Like CBC signals, long-duration signals from continuously emitting sources can also be lensed \cite{Suyamprakasam2025,Biesiada:2021pzo,Basak:2022fig,Liao:2019aqq, Sharma:2026lgx,Savastano:2022jjv}. Unlike a chirp which sweeps across a wide range of frequencies in a short span of time, CW sources are persistent and quasi-monochromatic \cite{Sieniawska:2019hmd,Haskell_Bejger2023}. The diffraction integral in Eq.~\eqref{eq:Fw} becomes 
\begin{equation} \label{eq:Fw(t)}
    F(t) = \frac{w}{2 \pi i} \int d^2 \bs{x}  e^{  i w T(\bs x,\bs y(t))}\,,
\end{equation}
being now a function of time through a time-varying impact parameter $y(t)$ due to relative motion of the source-lens system: 
\begin{equation}\label{eq:microlensing_y}
    y(t) = \sqrt{y_0^2 + \left( \frac{t-t_0}{T_E}\right)^2}\,,
\end{equation}
where $y_0$ is the impact parameter at closest approach between the lens and the source observed at time $t_0$. The quantity $T_E = R_E /v_{\perp}$ is the Einstein crossing time which defines the time taken by the source moving with a transverse velocity to cross the Einstein radius $R_E$ of the lens,  
\begin{equation}\label{eq:Einstein_crossing_time}
T_E = \frac{1}{|\bs{v}_{rel}|}\sqrt{\frac{4 G M_{L}}{c^2}D_L  \left(1-\frac{D_L}{D_S}\right)}\,,  
\end{equation}
where $\bs{v}_{rel}$ is the relative velocity ($|\bs{v}_{rel}| = v_\perp$, see Appendix B of  \cite{Kayser1986}), and $D_L$ and $D_S$ are the angular diameter distances to the lens and source, respectively. Note that within the Galaxy $z\approx 0$, the lens mass\footnote{We denote the lensing mass without the redshift term by $M_L$},  $M_{L}\approx M_{Lz}$ . Since the commonly-considered CW sources (rotating non-axisymmetric NSs) are located in the Galaxy we assumed $D_{LS} \approx D_S - D_L$. Equation~\eqref{eq:microlensing_y} assumes rectilinear relative motion and neglects the orbital motion of the Earth (microlensing parallax). The latter displaces the source trajectory in the lens plane by at most
$\pi_{\rm E}=D_{LS}/(R_{\rm E}D_S)\times\mathrm{AU}$ Einstein radii (e.g. for $M_L=100\,M_\odot$, $D_L=10$\,kpc, $D_S=20$\,kpc, $\pi_{\rm E}\simeq8\times10^{-3}$, which is negligible). For nearby, low-mass lenses the effect is larger (e.g.\ $\pi_{\rm E}\simeq0.3$ for $M_L=1\,M_\odot$, $D_L=1$\,kpc, $D_S=8$\,kpc) and the parallax term should be included in $y(t)$. Microlensing of CWs leads to a time-varying amplification curve analogous to the Paczy{\'n}ski curve in optical microlensing analysis \cite{Paczynski1986,Suyamprakasam2025}. This pattern can be recovered using standard CW search methods, e.g. the \texttt{TDFstat}\footnote{Time Domain $\mathcal{F}$-statistic method implementation: \url{https://polgraw.github.io/TDFstat/}} method, see Appendix~\ref{app:cw_lensing} for details. The search for such signals requires a long period of observations during which the Earth's motion has to be taken into account. The \texttt{gwlens\_cw\_datagen} command-line interface in {\gwlens} provides tools to generate lensed signals for CW sources in Gaussian noise with Doppler modulations and Solar System ephemerides. The data segments generated for a specific lens can be passed on to the \texttt{TDFstat} pipeline for lensed CW studies. In Fig.~\ref{fig:cwdata_pml_recovery} we show a six-day segment of lensed CW signal in Gaussian noise generated with \texttt{gwlens\_cw\_datagen}, with the signal's parameters chosen to match the P10 hardware injection \cite{GWOSC_CWHI,2017PhRvD..95f2002B}, but at the frequency $1000$ Hz, lensed by a $100\,M_{\odot}$ PML placed at $D_{L} = 10\,\mathrm{kpc}$, and a relative velocity of $200\,\mathrm{km\,s^{-1}}$. This gives an Einstein crossing time $T_E$ of $\sim 552$ days and requires $2T_E$ for a complete transit of the lens. Such a transit leads to a varying impact parameter and hence a time-dependent amplification of the signal over the observation period. The resulting GW microlensing pattern can be recovered with a semi-coherent $\mathcal{F}$ statistic analysis. Figure~\ref{fig:cwdata_pml_recovery} shows per-segment coherent $\mathcal{F}$-statistic SNRs, evaluated on six-day segments, matching the shape of the amplification pattern.  

\section{Conclusions}\label{sec:conclusion}

The increasing sensitivity of GW detectors and the growing catalogue of events make the detection of lensed GW signals increasingly likely. Interpretation of such signals in the wave-optics regime requires evaluating the diffraction integral for a variety of lens mass distributions. In this work we introduced the \textit{twisted-tail} method, a fast and accurate approach complementary to existing techniques. By identifying a point beyond the outermost image, we split the integration domain into a \textit{head} region, containing all images, which is integrated along the real axis with Gauss--Legendre quadrature, and a \textit{tail} region, which is rotated into the complex plane where the integrand decays exponentially and is integrated with Gauss--Laguerre quadrature.

The method is validated against the analytical point-mass solution, finding agreement to a relative error of ${\sim}10^{-13}$. Comparisons with {\glow} for more complicated lens models show good  agreement, see Figs.~\ref{fig:2dlens-comp} and \ref{fig:1dlens-comp}. The two codes have comparable cost for single evaluations, while {\gwlens} is faster when the amplification factor is required for many source positions at one or several frequencies. This combination of accuracy and efficiency is particularly relevant for non-trivial lenses, and for CW signals, for which the relative motion of source and lens with respect to the detector makes the amplification factor time-dependent, and requires its evaluation along the source trajectory. Future directions include using the method to search for lensed CBC signals, and coupling it with CW search pipelines for search and analysis of lensed CWs.

The implementation is publicly available in the Python package {\gwlens}. Although presented here for GW lensing, the  \textit{twisted-tail} technique applies to other fields involving similar integrals. These include femtolensing of gamma-ray bursts \cite{Ulmer:1994ij}, wave effects in the lensing of fast radio bursts \cite{Grillo:2018qjt}, and scintillation of radio sources \cite{Melrose2006}. By providing fast and accurate wave-optics amplification factors, {\gwlens} removes a computational bottleneck in the study of lensed GWs. It provides a practical tool for exploring dark-matter structure and other regimes of microlensing with current and future detectors.

\section*{Acknowledgements}
We thank Rico for useful comments and the internal LVK review of this manuscript. SH wishes to thank Apratim Ganguly for encouraging the pursuit of this research. SH is grateful to Margherita Grespan for her inputs on Levin and histogram-counting approaches to solving the diffraction integral. We gratefully acknowledge the Polish high-performance computing infrastructure PLGrid (ACK Cyfronet AGH) grant no. PLG/2025/018667, PLG/2026/019960, CAMK PAN computing resources, and the use of free and open software: \texttt{numpy} \cite{Harris_2020}, \texttt{LALSuite} \cite{lalsuite}, \texttt{scipy} \cite{Virtanen_2020}, \texttt{FINUFFT} \cite{Barnett2019_finufft,Barnett_2020_finufft}. This work was partially supported by the Polish National Science Centre (NCN) grants no. 2021/43/B/ST9/01714, 2023/49/B/ST9/02777 and 2025/56/C/ST9/00480. 

For the purpose of Open Access, the authors have applied a CC BY public copyright licence to any Author Accepted Manuscript (AAM) version arising from this submission. No real detector strain data were used in this work.

\appendix
\section{ Lensing potentials and library} \label{app: potentials and library}
Lensing potentials supplied by the user must be analytic and are passed to the \texttt{Potential1D} and \texttt{Potential2D} classes in the respective solver. These classes check if the functions are analytic using the complex-step method. {\gwlens} contains several built-in lensing potentials which are listed in Table \ref{tab:lens_summary}.

\begin{table*}[t]
  \caption{Lens potentials $\psi(x)$ available in \texttt{solver1d}. Any analytic $\psi(x)$ can be supplied in addition to these, and potentials can be added to build composite lenses. These potentials have identical forms in \texttt{solver2d}, with $x$ replaced by $|\boldsymbol{x}-\boldsymbol{x}_0|$ with $\bs{x}_0$ being their position. For non-axisymmetric lenses, e.g. external shear, $\psi$ depends on $(x_1,x_2)$ and not just on $|x|$.}
  \label{tab:lens_summary}
  \renewcommand{\arraystretch}{2.2}
  \setlength{\tabcolsep}{9.5pt}
  \begin{ruledtabular}
  \begin{tabular}{lllcc} 
    Model & \multicolumn{1}{c}{Parameters} & Function call & \multicolumn{1}{c}{$\psi(x)$} & \multicolumn{1}{c}{Reference} \\[0.8ex]
    \hline \hline
    Point mass      & $\psi_0$                & \texttt{solver1d.point\_mass} & $\psi_0\ln x$ & \cite{Keeton:2001ss} \\
    SIS             & $\psi_0$           & \texttt{solver1d.sis}         & $\psi_0\,x$ & \cite{Keeton:2001ss} \\
    CIS             & $\psi_0,\,x_c$     & \texttt{solver1d.cis}         & $\psi_0\!\left[\sqrt{x_c^2 + x^2} + x_c\ln\dfrac{2x_c}{\sqrt{x_c^2 + x^2}+x_c}\right]$ & \cite{Keeton:2001ss} \\
    NFW             & $\psi_0,\,x_s$       & \texttt{solver1d.nfw}         & $\dfrac{\psi_0}{2}\!\left[\ln^{2}\dfrac{x_s x}{2}+\arccos^{2}\dfrac{1}{x_s x}\right]$\,; $x_s = 
    {\xi_0}/{r_s}$ & \cite{Villarrubia-Rojo:2024xcj} \\
    {\renewcommand{\arraystretch}{1}%
     \begin{tabular}[c]{@{}l@{}}
       Plummer\\ sphere
     \end{tabular}} & $\psi_0,\,x_c$ & \texttt{solver1d.plummer}     & $\dfrac{1}{2}\,\psi_0\,x_c^{2}\ln\!\left(1+\dfrac{x^{2}}{x_c^{2}}\right)$ & \cite{Werner:2006dg} \\[1.5ex]
    {\renewcommand{\arraystretch}{1}%
     \begin{tabular}[c]{@{}l@{}}
       External\\ convergence\\ and shear
     \end{tabular}} & $\gamma_1,\,\gamma_2,\,\kappa$ & \texttt{solver2d.external\_shear} & $\dfrac{1}{2}\!\left[\kappa\,(x_1^{2}+x_2^{2}) + \gamma_1\,(x_1^{2}-x_2^{2}) + 2\gamma_2\,x_1x_2\right]$ & \cite{Mishra:2021xzz,Diego:2019lcd,Saha_2010} \\[1.5ex]
  \end{tabular}
  \end{ruledtabular}
\end{table*}

\section{CW lensing}
\label{app:cw_lensing} 

Lensed CWs differ qualitatively from lensed compact binary coalescences (CBCs) and, in general, short-duration transient signals. The most plausible sources, i.e. rotating non-axisymmetric NSs, as well as the lenses (stars, compact objects), are located inside the Galaxy, so the time delay between the lensed signals is shorter than the observation time: versions of the lensed signals are never resolved and overlap in the detector (the time delay is of the order of $4GM_{L}/c^3\simeq 2\times10^{-5}\,M_{L}/M_\odot$~s). The CW signals are nearly monochromatic, meaning that $w$ is fixed, but due to long duration of the signals, and possible relative transverse motion of source, lens and observer the impact parameter $y(t)$ changes in time. For a given source sky position $(\alpha,\delta)$, the amplitude modulation caused by the motion of the detector with respect to source has to be accounted for. In addition, the CW wavelength is comparable to the gravitational radius of the lens for some of the Galactic lenses, so that diffraction and interference effects are important. The $F(w,y(t))$ amplification therefore becomes a function of time, and the lensing manifests itself as a modulation of the amplitude and phase of an otherwise standard CW signal, in close analogy to the electromagnetic microlensing light curve \cite{Paczynski1986,OGLE1992,Suyamprakasam2025}. A practical consequence is that a sub-threshold constant-amplitude CW source may be temporarily magnified above the detection threshold, with the time-evolving pattern carrying information on the mass and kinematics of the lens.

{\gwlens} can be used to study lensed CW signals with the Time-Domain $\mathcal{F}$-statistic (\texttt{TDFstat}) implementation \cite{JKS98,Astone2010,tdfstat,Aasi_2014,PhysRevD.96.062002,PhysRevD.97.102003,PhysRevD.100.024004,PhysRevD.106.102008}. In the context of this work, the time-domain $\mathcal{F}$-statistic is a matched filter evaluated on narrow-band, uniformly sampled time series. Each series covers a segment whose length $T_0$ is an integer number of sidereal days, and a frequency band narrow enough that the noise spectral density $S_h$ can be treated as constant. The detector response is written as 
\begin{equation}
    h(t)=\sum_{i=1}^{4}A_i h_i(t)\,.
    \label{eq:cw_const_amp}
\end{equation}
The four constant amplitudes $A_i$ depend on the intrinsic amplitude $h_0$, source orientation angles $\iota$, $\psi_p$ and signal's constant phase $\phi_0$, while the basis functions $h_{1,3}=a(t)\{\cos,\sin\}\phi(t)$ and $h_{2,4}=b(t)\{\cos,\sin\}\phi(t)$ carry all the modulation caused by the motion of the detector. The amplitude modulation functions $a(t)$ and $b(t)$ reflect the influence of the antenna pattern: they are periodic with one sidereal day and depend on the location and orientation of the detector and on the sky position $(\alpha,\delta)$ of the source. 

Because the CW signals last for a long time, contrary to transient CBC signals (at least in the current detectors' frequency range), to coherently analyse long segments of data the movement of the detector with respect to the source has to be properly accounted for. A standard method is to alter the input data registered by the detectors located on the moving Earth, modifying it as if the detectors were at the Solar System Barycentre (SSB; the least-moving location in the Solar System). The signal phase is typically written as   
\begin{equation}
  \phi(t)=\sum_{k=0}^{s}\omega_k\frac{t^{k+1}}{(k+1)!}
  +\frac{\mathbf{n}_0\cdot\mathbf{r}_{\rm d}(t)}{c}
   \sum_{k=0}^{s}\omega_k\frac{t^{k}}{k!}\,,
   \label{eq:cw_phase}
\end{equation}
where $\omega_0$ and $\omega_k$ are the angular frequency and its time derivatives at the SSB at $t=0$, $\mathbf{n}_0$ is the unit vector towards the source, and $\mathbf{r}_{\rm d}(t)$ is the position of the detector relative to the SSB. The second term is the Doppler (R{\o}mer) modulation due to the orbital motion of the Earth and its daily rotation ($\mathbf{r}_{\rm d}$ is the sum of the barycentric position of the geocentre and the geocentric position of the detector). This phase correction is computed by the \texttt{gwlens\_cw\_datagen} routines using the JPL Solar System ephemerides \cite{JPL}, for the movement of the geocentre and an Earth-orientation model for the detector position \cite{Astone2010}. 

Maximizing the likelihood analytically over $A_i$ gives, for $T_0$ equal to an integer number of sidereal days \cite{JKS98}, a detection statistic 
\begin{eqnarray}
  2\mathcal{F} &\simeq& \frac{4}{S_h(f)\,T_0}
  \left(\frac{|F_a|^2}{\langle a^2\rangle}+\frac{|F_b|^2}{\langle b^2\rangle}\right)\,,\nonumber\\
  F_a &=& \int_0^{T_0}x(t)\,a(t)\,e^{-i\phi(t)}\,dt ,
  \label{eq:fstat}
\end{eqnarray}
and analogously for $F_b$, with $x(t)$ denoting the time-series input data, and $\langle\cdot\rangle$ the time average over the segment. In Gaussian noise the detection statistic $2\mathcal{F}$ follows a $\chi^2$ distribution with four degrees of freedom and non-centrality parameter equal to the squared optimal signal-to-noise ratio (SNR) $\rho^2\propto h_0^2T_0/S_h$, and the expectation value is $\mathrm{E}[2\mathcal{F}]=\rho^2 + 4$.

To evaluate $\mathcal{F}$ at every frequency of the band at once, the data are resampled to the SSB \cite{Schutz91,JKS98,Astone2010}. The signal phase can be written as $\phi(t)=\omega_0[t+\phi_{\rm m}(t)]+\phi_{\rm s}(t)$, where $\phi_{\rm m}=\mathbf{n}_0\cdot\mathbf{r}_{\rm d}/c$, and $\phi_{\rm s}$ denotes the spin-down terms. Introducing the barycentric time
$t_{\rm b}=t+\phi_{\rm m}(t)$ turns  Eq.~\eqref{eq:fstat} into  
\begin{equation}
  F_a\simeq\int_0^{T_0}x[t(t_{\rm b})]\,a[t(t_{\rm b})]\,
  e^{-i\phi_{\rm s}[t(t_{\rm b})]}\,e^{-i\omega_0 t_{\rm b}}\,dt_{\rm b},
\end{equation}
which allows an FFT in $t_{\rm b}$. Since $\phi_{\rm m}$ depends only on $(\alpha,\delta)$, the resampling is performed once per sky position. In the context of an all-sky search, the grid of signal parameter templates 
in $(f,\dot{f},\alpha,\delta)$ is constrained so that its nodes coincide with Fourier frequencies and steps in spin-down leave the sky position unchanged, so the resampling need not be repeated when only the spin-down changes \cite{Astone2010}. This procedure removes only the modulations caused by the motion of the detector. Any time-dependent lensing amplification, imprinted on the CW before it reaches the detector, survives in the resampled series as a slow variation of amplitude and phase in $t_{\rm b}$.

For the lensing amplification factor $F(t)=m(t)\,e^{i\Phi_{\rm L}(t)}$, the lensed version of signal in Eq.~\eqref{eq:cw_const_amp} is
\begin{equation}
  h_{L}(t)=m(t)\sum_{i=1}^{4}A_i\,h_i(t)\Big|_{\phi\to\phi-\Phi_{\rm L}} ,
  \label{eq:cw_lensed_amp}
\end{equation}
with the same constant amplitudes $A_i$: the lens rescales the envelope of the signal by $m(t)$ and shifts its phase by $-\Phi_{\rm L}(t)$, while the detector response, $a(t)$ and $b(t)$, is unchanged. 

The amplification varies on the Einstein crossing time $T_{\rm E}$ modified by $w$; for $w\gtrsim1$, the timescale is $T_{\rm f}\simeq T_{\rm E}/w$. As long as $T_0\ll\min(T_{\rm E},T_{\rm f})$, amplification $F$ is effectively constant within a segment and is absorbed into effective $h_0$ and $\phi_0$: Eq.~\eqref{eq:fstat} can be used unchanged, the recovered $(f,\dot f,\alpha,\delta)$ are unaffected, and the only imprint of lensing is a segment-dependent SNR, $\rho_k^2 = m^2(t_k)\,\rho_{0,k}^2$ (the ``Paczy{\'n}ski curve''), with $t_k$ the mid-time of the $k$-th segment, $\rho_{0,k}$ the unlensed value, accounting for the noise level and antenna response of that segment, as demonstrated on examples in \cite{Suyamprakasam2025}.

\bibliography{refs}
\end{document}